\pdfoutput=1
\documentclass[11pt,twoside,a4paper,cmspaper,final,collab]{cms-tdr}
\def\svnVersion{736d2b8-D}\def\svnDate{2024/07/23}\def\cmsCernNoTag{CERN-EP-2019-147}\def\cmsCernDate{\today}\def\cmsMessage{Published by the Journal of High Energy Physics as \href{http://dx.doi.org/10.1007/JHEP04(2020)171}{\doi{10.1007/JHEP04(2020)171}}\\[1em][including the erratum \href{http://dx.doi.org/10.1007/JHEP03(2022)187}{\doi{10.1007/JHEP03(2022)187}}]}
\begin{document}\cmsNoteHeader{HIG-17-027}

\RCS$Revision$
\RCS$HeadURL$
\RCS$Id$

\providecommand{\SUSHI} {\textsc{SusHi}\xspace}
\providecommand{\TwoHDMC} {\textsc{2hdmc}\xspace}
\providecommand{\TOPppTwo} {\textsc{Top++2.0}\xspace}

\newcommand{\mtt}{\ensuremath{m_{\ttbar}}\xspace}
\newcommand{\HH}{\ensuremath{\Phi}}
\newcommand{\HHOdd}{{\PSA}}
\newcommand{\HHEven}{{\PH}}
\newcommand{\thetaStar}{\ensuremath{\theta^*_{\cPqt_\ell}}}
\newcommand{\Irel}{\ensuremath{I_{\text{rel}}}}
\newcommand{\mtW}{\ensuremath{\mT^{\PW}}}

\newlength\cmsTabSkip\setlength{\cmsTabSkip}{1ex}

\cmsNoteHeader{HIG-17-027} \title{Search for heavy Higgs bosons decaying to a top quark pair in proton-proton collisions at $\sqrt{s} = 13$\TeV}

\date{\today}

\abstract{A search is presented for additional scalar (\HHEven) or pseudoscalar (\HHOdd) Higgs bosons decaying to a top quark pair in proton-proton collisions at a center-of-mass energy of 13\TeV. The data set analyzed corresponds to an integrated luminosity of 35.9\fbinv{} collected by the CMS experiment at the LHC. Final states with one or two charged leptons are considered. The invariant mass of the reconstructed top quark pair system and variables that are sensitive to the spin of the particles decaying into the top quark pair are used to search for signatures of the \HHEven{} or \HHOdd~bosons. The interference with the standard model top quark pair background is taken into account. A moderate signal-like deviation compatible with an \HHOdd~boson with a mass of 400\GeV{} is observed with a global significance of 1.9 standard deviations. New stringent constraints are reported on the strength of the coupling of the hypothetical bosons to the top quark, with the mass of the bosons ranging from 400 to 750\GeV and their total relative width from 0.5 to 25\%. The results of the search are also interpreted in a minimal supersymmetric standard model scenario. Values of $m_\HHOdd$ from 400 to 700\GeV are probed, and a region with values of $\tan\beta$ below 1.0 to 1.5, depending on $m_\HHOdd$, is excluded at 95\% confidence level.}

\hypersetup{pdfauthor={CMS Collaboration},pdftitle={Search for heavy Higgs bosons decaying to a top quark pair in proton-proton collisions at sqrt(s) = 13 TeV},pdfsubject={CMS},pdfkeywords={CMS, physics, exotica, top, higgs, 2HDM, MSSM}}

\maketitle

\section{Introduction}
\label{sec:Introduction}
The observation of a Higgs boson with a mass of 125\GeV by the ATLAS and CMS experiments~\cite{Aad:2012tfa,Chatrchyan:2012xdj,Chatrchyan:2013lba}
was a milestone in particle physics, confirming the existence of a crucial ingredient of the standard model (SM) of particle physics.
Multiple extensions of the SM predict new spin-0 states. These include two-Higgs-doublet models (2HDMs)~\cite{Branco:2011iw}, of which the minimal supersymmetric standard model
(MSSM)~\cite{Wess:1974tw,Dimopoulos:1981zb} is a particular realization, models predicting a new electroweak singlet~\cite{Huitu:2019}, and other models with a combination of singlet and doublet fields~\cite{Muhlleitner:2017dkd}.
The additional bosons may also provide a portal to dark matter, by acting as a mediator between SM and dark matter particles~\cite{DMLHC:2015,Arina:2016}.

The new states introduced in these extensions of the SM may include charged Higgs bosons, \Hpm, scalar (CP-even) neutral \HHEven{} and \Ph{} bosons (here \Ph{} denotes the lighter of the two states), and a pseudoscalar (CP-odd) neutral \HHOdd{} boson.
For convenience and depending on the context, a common symbol \HH{} is used in this paper to represent the \HHEven{} and \HHOdd~bosons.

Top quarks play a key role in searches for new physics because of their high mass and large coupling to the SM Higgs boson.
Provided that additional Higgs bosons couple to fermions via a Yukawa interaction, the top quark's high mass suggests the size of the coupling to these new bosons to be large as well.
Hence, assuming the masses of the new \HH~bosons are sufficiently high, their possible decay to a top quark pair is particularly interesting.
Decays of CP-odd Higgs bosons to weak vector bosons, $\HHOdd \to \PW\PW$ and $\HHOdd \to \PZ\PZ$, are forbidden (at tree level) if the CP symmetry is assumed.
Such decays are also strongly suppressed for \HHEven~bosons in the vicinity of the alignment limit of 2HDMs,
in which the properties of the \Ph~boson approach those of the SM Higgs boson~\cite{Craig:2013hca}. 
The aligned scenario may naturally occur as a result of a broken symmetry, and such models imply the existence of additional Higgs bosons at relatively low mass ($\lesssim 550$\GeV)~\cite{Lane:2019}.
In this paper, however, we do not rely on the assumption of alignment and its naturalness, and probe masses between 400 and 750\GeV.

We consider a Yukawa-like coupling between the new spin-0 bosons and the top quark.
The corresponding terms in the Lagrangian for the two CP eigenstates read as
\begin{linenomath}
\begin{equation}
 \label{Eq:coupling}
 \mathcal{L}_\text{Yukawa,\,\HHEven} = -g_{\HHEven\ttbar}\,\frac{m_{\cPqt}}{v}\,\cPaqt\cPqt\HHEven,
      \qquad \mathcal{L}_\text{Yukawa,\,\HHOdd} = ig_{\HHOdd\ttbar}\,\frac{m_{\cPqt}}{v}\,\cPaqt{\gamma}_{5}\cPqt\HHOdd,
\end{equation}
\end{linenomath}
where $m_{\cPqt}$ is the top quark mass, $v$ is the SM Higgs vacuum expectation value, and the strength of the couplings is controlled by real-valued coupling modifiers $g_{\HH\ttbar} \geqslant 0$.

A special case of a Type-II 2HDM~\cite{Branco:2011iw} is the Higgs sector in the hMSSM~\cite{Djouadi:2013uqa}, where the \Ph~boson is identified with the Higgs boson with mass of 125\GeV.
The hMSSM can be fully described by two tree-level parameters: $\tan\beta$, the ratio of the vacuum expectation values of the two Higgs fields,
and $m_\HHOdd$, the mass of the pseudoscalar boson.
The parameter region at low values of $\tan\beta$ is of particular interest, since the coupling of the additional Higgs bosons to top quarks is enhanced in this regime.

\begin{figure}
  \centering
\includegraphics[height=15ex]{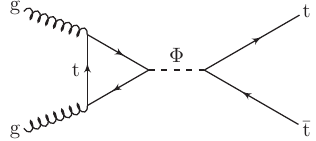}\hspace{5ex}
\includegraphics[height=15ex]{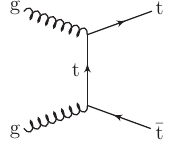}
\caption{The Feynman diagram for the signal process (left) and an example diagram for the SM production of top quark pairs (right).}
\label{fig:feynman}
\end{figure}

We consider the production of new \HH~bosons through the gluon fusion process, with only top quarks in the loop.
When the heavy Higgs boson decays into a top quark pair, this mode interferes at the quantum level with the SM production of top quark pairs.
Example Feynman diagrams are shown in Fig.~\ref{fig:feynman}.
As a consequence, the signal consists of a resonant and an interference component.
The resonant component corresponds to the square of the amplitude given by the signal diagram, and results in a Breit--Wigner peak
in the distribution of the invariant mass of the \ttbar~system, \mtt.
The interference component may be either destructive or constructive, depending on the phase space region and signal model.
The sum of the components may result in a peak-dip structure in the \mtt~distribution~\cite{Gaemers:1984sj,Dicus:1994bm,Bernreuther:1997gs}.
It is worth noting that the shape and magnitude of the interference depends on the specific signal model, 
and can be significantly modified by new particles appearing in the loop of the production diagram~\cite{Carena:2016,Djouadi:2019}. 

Decays of the scalar and pseudoscalar Higgs bosons produce \ttbar~pairs in the ${}^3P_0$ and ${}^1S_0$ states respectively~\cite{Bernreuther:1997gs}, while the SM $gg \to \ttbar$ production results in a mixture of states that changes with the partonic center-of-mass energy.
Consequently, the signal and the background exhibit different angular properties, providing an additional handle to distinguish them.

A search for \HHEven{} or \HHOdd~bosons decaying to a top quark pair was performed at a center-of-mass energy $\sqrt{s} = 8$\TeV by the ATLAS experiment~\cite{atlashttottbar8TeV}.
The results were interpreted within the context of a Type-II 2HDM.
The CMS experiment performed a search for top quark associated production of an \HHEven{} or \HHOdd~boson decaying to a top quark pair at $\sqrt{s} = 13$\TeV~\cite{cmsSUSYsamesign13TeV}.
The ATLAS and CMS Collaborations also searched for spin-1 and spin-2 resonances decaying to a top quark pair~\cite{atlasZprime13TeV,cmsZprime13TeV},
generally probing very high masses and Lorentz-boosted topologies, and without considering quantum interference with SM top quark pair production.
In addition, both collaborations performed searches for \Hpm{} decaying to a top and a bottom quark~\cite{Aaboud:2018cwk, Khachatryan:2015qxa}, which are also sensitive to the region of low $\tan\beta$ in the hMSSM parameter space.

This paper describes a search for \HHEven{} or \HHOdd~bosons decaying to a top quark pair in proton-proton
collisions at $\sqrt{s} = 13$\TeV using the CMS detector at the CERN LHC. The data set analyzed corresponds to an integrated luminosity of about 35.9\fbinv, collected in 2016.
Events are selected in which the top quark pair decays into a final state with one or two leptons, where a lepton refers to an electron or a muon throughout this article.
This analysis exploits both \mtt and angular variables sensitive to the spin of the heavy Higgs bosons.
Constraints on the coupling modifier $g_{\HH\ttbar}$ are derived as a function of the boson mass and width.
The results are also interpreted in the hMSSM context, putting constraints in the $(m_\HHOdd, \tan\beta)$ plane.

\section{The CMS detector and event reconstruction}
\label{sec:detector}
The central feature of the CMS apparatus is a superconducting solenoid of 6\unit{m} internal diameter, providing a magnetic field of 3.8\unit{T}.
Within the solenoid volume are a silicon pixel and strip tracker, a lead tungstate crystal electromagnetic calorimeter (ECAL),
and a brass and scintillator hadron calorimeter (HCAL), each composed of a barrel and two endcap sections.
Forward calorimeters extend the pseudorapidity coverage provided by the barrel and endcap detectors.
Muons are detected in gas-ionization chambers embedded in the steel flux-return yoke outside the solenoid.
A more detailed description of the CMS detector, together with a definition of the coordinate system
used and the relevant kinematic variables, can be found in Ref.~\cite{Chatrchyan:2008zzk}.

The reconstructed vertex with the largest value of summed physics-object $\pt^2$ is taken to be the primary $\Pp\Pp$ interaction vertex.
The physics objects are the jets, clustered using the jet finding algorithm~\cite{Cacciari:2008gp,Cacciari:2011ma} with the tracks assigned to the vertex as inputs, and the associated missing transverse momentum, taken as the negative vector sum of the \pt of those jets.

The particle-flow (PF) algorithm~\cite{CMS-PRF-14-001} aims to reconstruct and identify each individual particle in an event, with an optimized combination of information from the various elements of the CMS detector.
The reconstructed particles are referred to as PF~candidates in the following.
The energy of photons is obtained from the ECAL measurement.
The energy of electrons is determined from a combination of the electron momentum at the primary interaction vertex as determined by the tracker, the energy of the corresponding ECAL cluster, and the energy sum of all bremsstrahlung photons spatially compatible with originating from the electron track.
The energy of muons is obtained from the curvature of the corresponding track.
The energy of charged hadrons is determined from a combination of their momentum measured in the tracker and the matching ECAL and HCAL energy deposits, corrected for zero-suppression effects and for the response function of the calorimeters to hadronic showers.
Finally, the energy of neutral hadrons is obtained from the corresponding corrected ECAL and HCAL energies.

Jets are reconstructed from the PF candidates with the infrared and collinear safe anti-\kt algorithm~\cite{Cacciari:2011ma} operated with a distance parameter $R = 0.4$.
Jet momentum is determined as the vectorial sum of all particle momenta in the jet, and is found from simulation to be, on average, within 5 to 10\% of the true momentum over the whole \pt~spectrum and detector acceptance.
Additional $\Pp\Pp$~interactions within the same or nearby bunch crossings, referred to as pileup, can contribute additional tracks and calorimetric energy depositions to the jet momentum.
To mitigate this effect, tracks identified to be originating from pileup vertices are discarded, and an offset correction is applied to correct for remaining contributions.
Jet momentum corrections are derived from simulation to bring the measured response of jets to that of particle level jets on average.
In situ measurements of the momentum balance in dijet, $\text{photon} + \text{jet}$, $\PZ + \text{jet}$, and quantum chromodynamics (QCD) multijet events
(consisting uniquely of jets produced through the strong interaction),
are used to account for any residual differences in jet \pt~scale in data and simulation~\cite{Khachatryan:2016kdb}.
The relative jet \pt{} resolution
amounts typically to 15--20\% at 30\GeV, 10\% at 100\GeV, and 5\% at 1\TeV.
Additional selection criteria are applied to each jet to remove jets potentially dominated by anomalous contributions from various subdetector components or reconstruction failures~\cite{PAS-JME-16-003}.

Jets originating from \cPqb~quarks are identified with the cMVAv2 algorithm~\cite{CMS-BTV-16-002},
combining six different \cPqb~jet discriminators, which exploit displaced track and secondary vertex information.
The collection of \cPqb-tagged jets is defined by an operating point that corresponds to a \cPqb~tagging efficiency of about 66\% for \cPqb~jets, 13\% for \cPqc~jets, and a misidentification probability (``mistag rate'') of about 1\% for light-flavor jets.
Differences between data and simulation in the \cPqb~tagging efficiency and mistag rate are accounted for by scale factors that depend on the jet \pt{} and $\eta$.

The electron momentum is estimated by combining the energy measurement in the ECAL with the momentum measurement in the tracker.
The momentum resolution for electrons with $\pt \approx 45\GeV$ from $\PZ \to \Pe \Pe$ decays ranges from 1.7\%
for nonshowering electrons in the barrel region to 4.5\% for showering electrons in the endcaps~\cite{Khachatryan:2015hwa}.
Muons are measured in the pseudorapidity range $\abs{\eta} < 2.4$, with detection planes made using three technologies: drift tubes, cathode strip chambers, and resistive plate chambers.
Matching muons to tracks measured in the silicon tracker results in a relative transverse momentum resolution for muons with $20 <\pt < 100\GeV$ of 1.3--2.0\% in the barrel and better than 6\% in the endcaps.
The \pt~resolution in the barrel is better than 10\% for muons with \pt up to 1\TeV~\cite{Sirunyan:2018fpa}.
Simulation-to-data scale factors that depend on the lepton \pt{} and $\eta$ are used to correct for small differences in lepton trigger, identification, and isolation efficiency.

We define tight and loose collections of electron and muon candidates, corresponding to the stringency of the lepton identification criteria.
For electrons, an updated version of the criteria from Ref.~\cite{Khachatryan:2015hwa} is utilized, while the muon identification is as described in Ref.~\cite{Sirunyan:2018fpa}.
Tight and loose electrons are furthermore required to satisfy $\pt > 20$\GeV, while tight (loose) muons have $\pt > 20~(10)$\GeV.
The relative lepton isolation, \Irel, is calculated as the sum of the transverse momenta of charged-hadron,
neutral-hadron, and photon PF candidates, inside a cone of $\Delta R = \sqrt{\smash[b]{(\Delta\eta)^2 + (\Delta\phi)^2}} = 0.4$ around the lepton, divided by the lepton \pt.
An estimated contribution from pileup is subtracted in this calculation.
Tight electrons must satisfy $\Irel \lesssim 0.06$, while loose ones are required to have $\Irel \lesssim 0.18$ (0.16) in the barrel (endcap) region.
Tight (loose) muon candidates must satisfy $\Irel < 0.15$ (0.25).

The variable \ptmiss, referred to as the missing transverse momentum~\cite{Sirunyan:2019kia}, is defined as the magnitude of the missing transverse momentum vector \ptvecmiss,
which is the projection on the plane perpendicular to the beams of the negative vector sum of the momenta of all reconstructed PF~candidates in the event.
The energy corrections applied to the jets are propagated to the \ptmiss{} calculation.

\section{Data and simulated event samples}
\label{sec:samples}
This analysis is performed on a $\Pp\Pp$ collision data set recorded during 2016, at a center-of-mass energy of 13\TeV.
The total integrated luminosity of the collected data sample is $35.9 \pm 0.9$\fbinv~\cite{CMS-PAS-LUM-17-001}.

The single-electron (single-muon) data sample is selected with triggers~\cite{Khachatryan:2016bia} based on the presence of an isolated electron (muon).
The dielectron, electron-muon, and dimuon data samples are selected with triggers that require the presence of two leptons of the corresponding flavors.
In order to increase the selection efficiency for dilepton events in which the subleading lepton has a relatively low \pt, all the dilepton samples are further extended with events that pass the single-lepton but not the dilepton triggers.

In order to compare the collected data to theoretical predictions, Monte Carlo (MC) samples are produced with events simulating the $\HH \to \ttbar$ signal and SM background processes.
The signal is simulated at leading order (LO) accuracy in perturbative QCD using a custom model in the \MGvATNLO 2.5.1 event generator~\cite{Alwall:2014hca}
that implements the top quark loop of the gluon fusion production via an effective coupling between the new boson and gluons~\cite{Spira:1995rr}.
The generator employs the NNPDF3.0 parton distribution functions (PDFs)~\cite{Ball:2014uwa},
and is interfaced with \PYTHIA~8.212~\cite{Sjostrand:2014zea} for fragmentation and hadronization, with the CUETP8M1 underlying event tune~\cite{Skands:2014pea,Khachatryan:2015pea}.
Signal event samples are produced for (pseudo)scalar boson masses of 400, 500, 600, and 750\GeV, with relative total decay widths $\Gamma_\HH / m_\HH$ of 2.5, 5, 10, and 25\% for each mass scenario.
The factorization and renormalization scales, $\mu_\text{F}$ and $\mu_\text{R}$, are set on an event-by-event basis to $\mtt / 2$,
following the choice in Ref.~\cite{Hespel:2016qaf}. The top quarks from the heavy Higgs boson decay are decayed in \MGvATNLO, preserving their spin correlations.
Samples are generated for events corresponding to the resonant heavy boson signal,
and for events corresponding to interference terms in the matrix element calculation between the signal and SM \ttbar{} background.
Events in the interference samples can receive negative weights, which reflects the sign of the corresponding part of the squared matrix element in the presence of a destructive interference.
Since the heavy Higgs boson is produced via gluon fusion with a top quark loop, the coupling between the boson and the top quark appears twice in the matrix element.
As a result, events originating from the resonance (interference) matrix element terms correspond to a cross section that is proportional to $g^4_{\HH\ttbar}$ ($g^2_{\HH\ttbar}$).

We calculate the next-to-next-to-LO (NNLO) cross sections for the resonant part of a given signal using the \SUSHI~1.6.1 program~\cite{Harlander:2012pb}.
The ratio of the NNLO cross section over the LO cross section computed with \MGvATNLO determines the $K$~factor, typically of size ${\approx}2$, applied to the resonant part of the signal.
The $K$~factors applied to the interference component of the signal are obtained as the geometric mean of the $K$~factors of the resonant signal and the SM \ttbar{} process~\cite{Hespel:2016qaf}.
The SM~\ttbar{} $K$~factor is 1.6, calculated as the ratio between the NNLO cross section used for the simulated \ttbar~sample,
as described below, and the LO cross section obtained in a similar setup.
The $K$~factors for the resonant part of the signal and the interference are applied throughout this analysis.
In the hMSSM interpretation, we also use the \TwoHDMC program~\cite{Eriksson:2009ws} to calculate, for given $m_\HHOdd$ and $\tan\beta$, the mass of the \HHEven~boson and the widths of both heavy Higgs bosons, as well as other MSSM parameters.

The main SM background contribution in this analysis originates from \ttbar~production.
Other background events originate from single top quark production, single boson production (Drell--Yan $\PZ/\gamma^* + \text{jets}$ and $\PW + \text{jets}$),
diboson processes ($\PW\PW$, $\PW\PZ$, and $\PZ\PZ$), \ttbar~production in association with a \PZ{} or \PW~boson (commonly referred to as $\ttbar \PV$), and QCD multijet processes.

The \ttbar~process is simulated to next-to-LO (NLO) using the \POWHEG~v2 generator~\cite{Nason:2004rx,Frixione:2007vw,Alioli:2010xd,Campbell:2014kua}, assuming a top quark mass of 172.5\GeV.
The factorization and renormalization scales are set to $\sqrt{\smash[b]{m_{\cPqt}^2 + p_{\text{T},\cPqt}^2}}$, where $m_{\cPqt}$ and $p_{\text{T},\cPqt}$ are the mass and the transverse momentum of the top quarks in the underlying Born-level configuration.
The NNPDF3.0 PDF set is used, and the events are passed to \PYTHIA with the CUETP8M2T4 event tune~\cite{CMS-PAS-TOP-16-021}.
The predicted \ttbar~production cross section is $831.8^{+19.8}_{-29.2}$ (scale) ${}\pm 35.1$ (PDF + \alpS)~pb,
as calculated with the \TOPppTwo program to NNLO in perturbative QCD, including soft-gluon resummation to next-to-next-to-leading logarithmic (NNLL) order (as discussed in Ref.~\cite{Czakon:2011xx} and references therein),
and assuming a top quark mass of 172.5 GeV. The first uncertainty comes from the independent variation of the factorization and renormalization scales,
while the second one is associated to variations in the PDF and strong coupling \alpS,
following the PDF4LHC prescription with the MSTW2008 68\% confidence level NNLO, CT10 NNLO, and NNPDF2.3 5f FFN PDF sets (as discussed in Ref.~\cite{Botje:2011sn} and references therein, and Refs.~\cite{Martin:2009bu,Gao:2013xoa,Ball:2013}).
The modeling of SM \ttbar{} production in \POWHEG is known to predict a harder \pt~spectrum of the top quarks than observed in the data. An empirical reweighting for top quark pairs based on the \pt~spectrum of generator-level top quarks is applied to obtain
a better agreement with the measured differential \ttbar{} cross section~\cite{Khachatryan:2016mnb,differentialXSdilep13TeV}.

The single top quark production processes, via the $t$, $\PQt\PW$, and $s$ channels, are generated at NLO using \POWHEG~v2, \POWHEG, and \MGvATNLO, respectively.
The samples are normalized using the NLO cross section predictions for the $t$ and $s$~channels~\cite{singletopNLO1,singletopNLO2},
and approximate NNLO prediction for the $\PQt\PW$ channel~\cite{Kidonakis:2013zqa}.
The $\PZ/\gamma^* + \text{jets}$, $\PW + \text{jets}$, and $\ttbar \PV$ event samples are generated using \MGvATNLO. For the single boson production processes,
events with up to four additional partons are generated at LO, and the MLM matching scheme~\cite{Alwall:2007fs} is employed to combine the different parton multiplicities.
The single-boson production cross sections are calculated at NNLO~\cite{Melnikov:2006kv,fewz2}. In the dilepton analysis, the normalization of the $\PZ/\gamma^* + \text{jets}$ contribution is determined from a control region in data.
The $\ttbar \PV$ events are generated at NLO, applying \MCATNLO~\cite{Frixione:2002ik} merging, and are normalized using NLO cross section predictions.
Events simulating the diboson processes are generated using \PYTHIA and normalized to the respective NNLO (for WW production)~\cite{wwNNLO} or NLO (for WZ and ZZ production)~\cite{mcfm} cross sections.
The modeling of QCD multijet events is obtained from a control region in data, but events simulated with \PYTHIA are used to validate the modeling.
All events are interfaced with \PYTHIA for the fragmentation and hadronization, using the CUETP8M2T4 (single top processes) or CUETP8M1 (others) tunes.

The simulated events are processed through the CMS detector simulation based on the \GEANTfour program~\cite{Agostinelli:2002hh}.
Pileup events generated with \PYTHIA are overlaid in all samples, to simulate additional interactions in the same bunch crossing.
The simulated events are weighted to reproduce the distribution of the number of pileup interactions observed in data.
On average, there are 23 collisions per bunch crossing.

\section{Data analysis}
\label{sec:analysis}
We search for heavy Higgs bosons decaying into a top quark pair, in final states with either one or two leptons, where the lepton is either an electron or a muon.
The analysis strategy for single-lepton final states differs from the one for dilepton final states,
due to differences in the event selection, SM background composition, kinematic top quark pair reconstruction, and definition of observables that discriminate between the SM background and the signal.

\subsection{Single-lepton final state}
\label{sec:singlelepanalysis}
In the single-lepton channel we aim to select events originating from top quark pair decays to a leptonically decaying \PW~boson and a hadronically decaying \PW~boson.
The targeted topology is therefore $\PQt\PAQt \to \ell^+ \PGn \PQb ~ \PQq\PAQq' \PAQb$,
where $\ell$ denotes an electron or a muon and the leptonic and hadronic \PW~boson decays may be swapped.
Events in the single-electron (single-muon) channel in data and simulation are required to pass a single-electron (single-muon) trigger, as explained in Section~\ref{sec:samples}.
Selected events must have exactly one tight electron (muon) with $\pt > 30~(26)$\GeV and $\abs{\eta_{\text{SC}}} < 2.5$ ($\abs{\eta} < 2.4$), where $\eta_\text{SC}$ is the pseudorapidity of the ECAL supercluster associated with the electron~\cite{Khachatryan:2015hwa}.
To suppress the contribution from $\PZ/\gamma^* + \text{jets}$ and other processes in which multiple prompt leptons are produced,
an event is rejected if an additional loose electron or loose muon is found.
This also ensures orthogonality to the event selection of the dilepton analysis outlined in Section~\ref{sec:dilepanalysis}.
An event must contain at least four jets with $\pt > 20$\GeV and $\abs{\eta} < 2.4$, at least two of which are required to be \cPqb~tagged.
To further suppress the QCD multijet background, only events with $\mtW > 50$\GeV are selected.
The transverse mass variable is defined as $\mtW = \sqrt{\smash[b]{2\pt^\ell\ptmiss [1 - \cos\Delta\phi(\ptvec^\ell, \ptvecmiss)]}}$,
with $\ptvec^\ell$ being the transverse momentum of the only tight electron or muon in the event.

Each event that passes the selection described above is reconstructed under the assumption that it has been produced in the process $\PQt\PAQt \to \ell\PGn\PQb\PAQb\PQq\PAQq'$.
The \ttbar~system is reconstructed using an approach similar to the one adopted in Ref.~\cite{Khachatryan:2016mnb}.
All possible ways to assign four reconstructed jets to the four quarks in the final state are considered.
To reduce the number of combinations, each of the two \cPqb~quarks is required to be associated to a \cPqb-tagged jet.
For each considered choice of a jet assigned to a \cPqb~quark stemming from the semileptonically decaying top quark, an attempt is made to reconstruct the transverse momentum of the neutrino, $\ptvec^{\PGn}$~\cite{Betchart:2013nba}.
This is achieved by imposing a constraint on the mass of the semileptonically decaying top quark and on the mass of the leptonically decaying \PW~boson,
and choosing the unique solution of the neutrino momentum that minimizes the distance $D_{\PGn} = \abs{\ptvec^{\PGn} - \ptvecmiss}$.
Next, a likelihood function is constructed from the probability density function of the minimal value of $D_{\PGn}$,
and the two-dimensional (2D) probability density function of the reconstructed mass of the top quark and the \PW~boson in the hadronic side of the \ttbar~decay.
The jet assignment with the largest value of this likelihood is chosen to reconstruct the \ttbar~system.

The performance of the \ttbar~reconstruction algorithm is studied using SM~\ttbar simulation, considering only events with the targeted decays at the generator level, $\PQt\PAQt \to \ell\PGn\PQb\PAQb\PQq\PAQq'$.
A correct jet-quark assignment exists for 44\% of such events that pass the event selection.
For those events where a correct assignment exists and a solution can be found for the neutrino momentum, the probability that all four jets are correctly assigned to the quarks
varies from around 60 to 80\%, depending on the value of the invariant mass of the generator-level top quark pair, $\mtt^\text{gen}$.
The relative \mtt~resolution, as computed with all selected events with targeted decays, changes from about 17 to 21\%, depending on $\mtt^\text{gen}$.

\begin{figure}[h]
  \centering
  \includegraphics[width=0.5\textwidth]{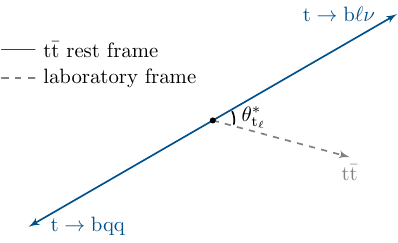}
  \caption{Definition of angle~$\thetaStar$ used in the single-lepton final state. Momenta in different coordinate systems are depicted with arrows of different colors and styles.}
  \label{fig:ljets_angle}
\end{figure}

The \ttbar~reconstruction results in a solution in about 85\% of observed events, and only the events with a solution are considered for further analysis.
The search for the $\HH \to \ttbar$ signal is performed using two observables.
The first one is the invariant mass~\mtt, as obtained from the \ttbar~reconstruction algorithm, and probes the mass of the heavy (pseudo)scalar boson.
The second observable is $\abs{\cos\thetaStar}$, where $\thetaStar$ denotes the angle between the momentum of the semileptonically decaying top quark in the
\ttbar~rest frame and the momentum of the \ttbar~system in the laboratory frame, as illustrated in Fig.~\ref{fig:ljets_angle}.
In a $\HH \to \ttbar$ signal process, the heavy Higgs boson would decay into top quarks isotropically, resulting in a flat distribution of $\cos\thetaStar$ at the generator level.
The SM \ttbar{} production, on the other hand, yields a distribution peaking at $\pm 1$.
As a result, the distribution of $\abs{\cos\thetaStar}$ is relatively enriched in signal events towards $\abs{\cos\thetaStar} = 0$.

The QCD multijet background is estimated from dedicated control regions in the data, independently in the electron and muon channels.
The total event yield for this background is obtained from data using a variant of the ABCD method~\cite{Loginov:2010zz}.
Four regions are defined, based on the relative isolation of the lepton (smaller or greater than the isolation threshold imposed on the tight lepton),
and the \mtW{} variable (smaller or greater than 50\GeV).
The three regions complementary to the signal region are relatively enriched in the multijet background.
The overall rate of QCD multijet background in the signal region is estimated with a simultaneous fit to the numbers of events observed in the four regions,
exploiting the factorization of the distribution of this background in $(\mtW, I_{\text{rel}})$.
The shape of the multijet distribution of the observables \mtt and $\abs{\cos\thetaStar}$ is determined
from data using events with an inverted lepton isolation selection applied, after subtracting the contributions from other backgrounds.

The data and the expected SM background yields are shown in Table~\ref{tab:eventyields_singlelep}, for selected events that have a solution of the \ttbar~reconstruction algorithm.
The background predictions are computed with the help of a maximum likelihood fit to the data using the background-only version of the full statistical model that will be described in Section~\ref{sec:Results}.
The uncertainties obtained in this fit, referred to as post-fit uncertainties, are reported.

\begin{table}[ht]
\topcaption{
Event yields and composition of SM background in the single-electron and single-muon channels.
Expected yields are computed after the background-only fit to the data as explained in the text.
}
\centering
\begin{tabular}{l cc }
\hline
\multirow{2}{*}{Process}  & Electron channel  & Muon channel  \\
 & \multicolumn{2}{c}{Event yield}  \\
\hline
Observed                      & 274\,821  & 416\,254  \\
Total background              & $\left(274.8^{+0.8}_{-0.9}\right)\ten{3}$  & $\left(416.3^{+1.1}_{-1.2}\right)\ten{3}$  \medskip\\
\hline
 & \multicolumn{2}{c}{Fraction w.\,r.\,t. total background}  \\
\hline
\ttbar                        & 91.9\%  & 92.1\% \\
Single top quark              & 3.9\%   & 4.0\%  \\
$\PW + \text{jets}$           & 1.9\%   & 2.1\%  \\
$\PZ/\gamma^* + \text{jets}$  & 0.4\%   & 0.3\%  \\
$\ttbar \PV$                  & 0.2\%   & 0.2\%  \\
Diboson                       & 0.1\%   & 0.1\%  \\
QCD multijet                  & 1.5\%   & 1.0\%  \\
\hline
\end{tabular}
\label{tab:eventyields_singlelep}
\end{table}

\begin{figure}[p]
  \centering
  \includegraphics[width=\textwidth]{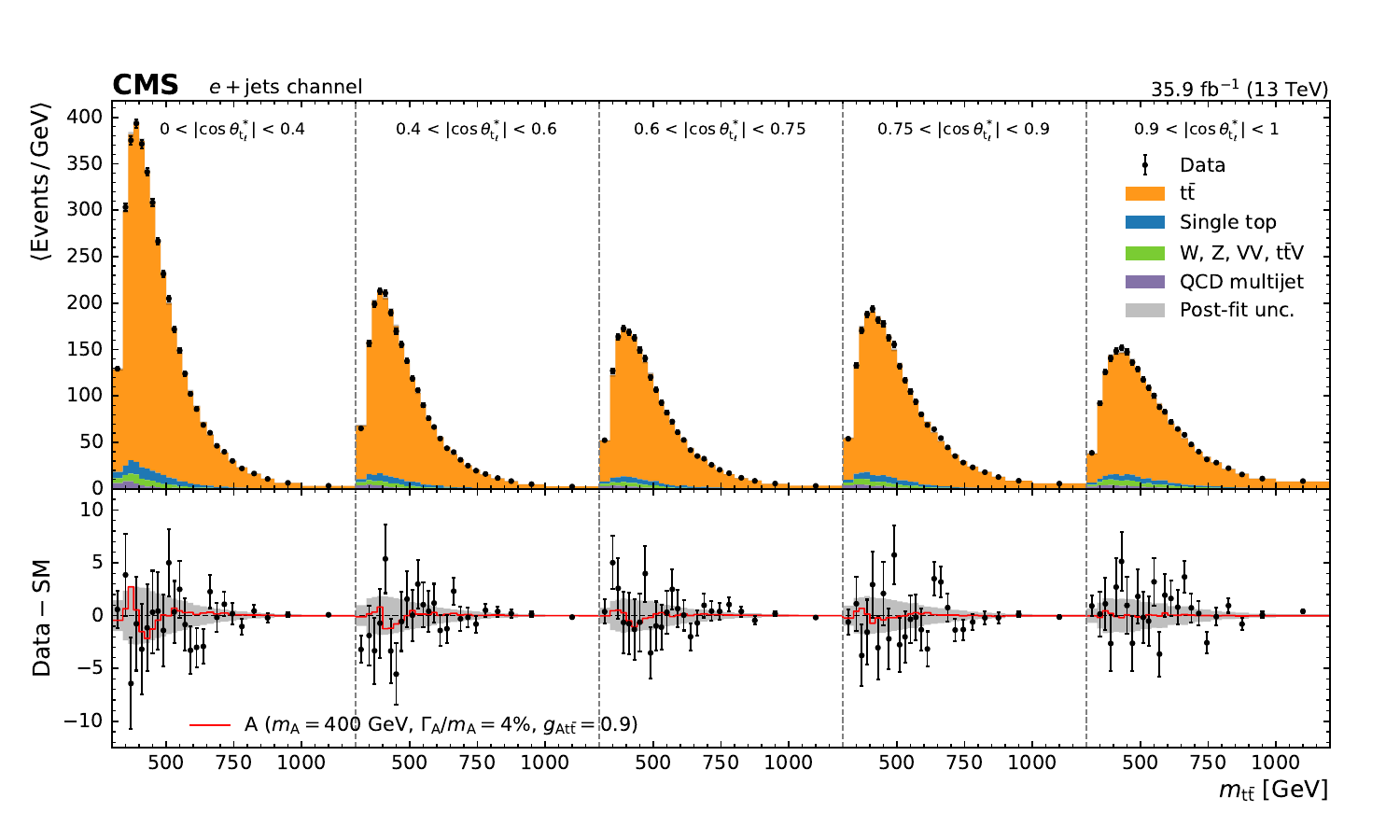} \\
  \includegraphics[width=\textwidth]{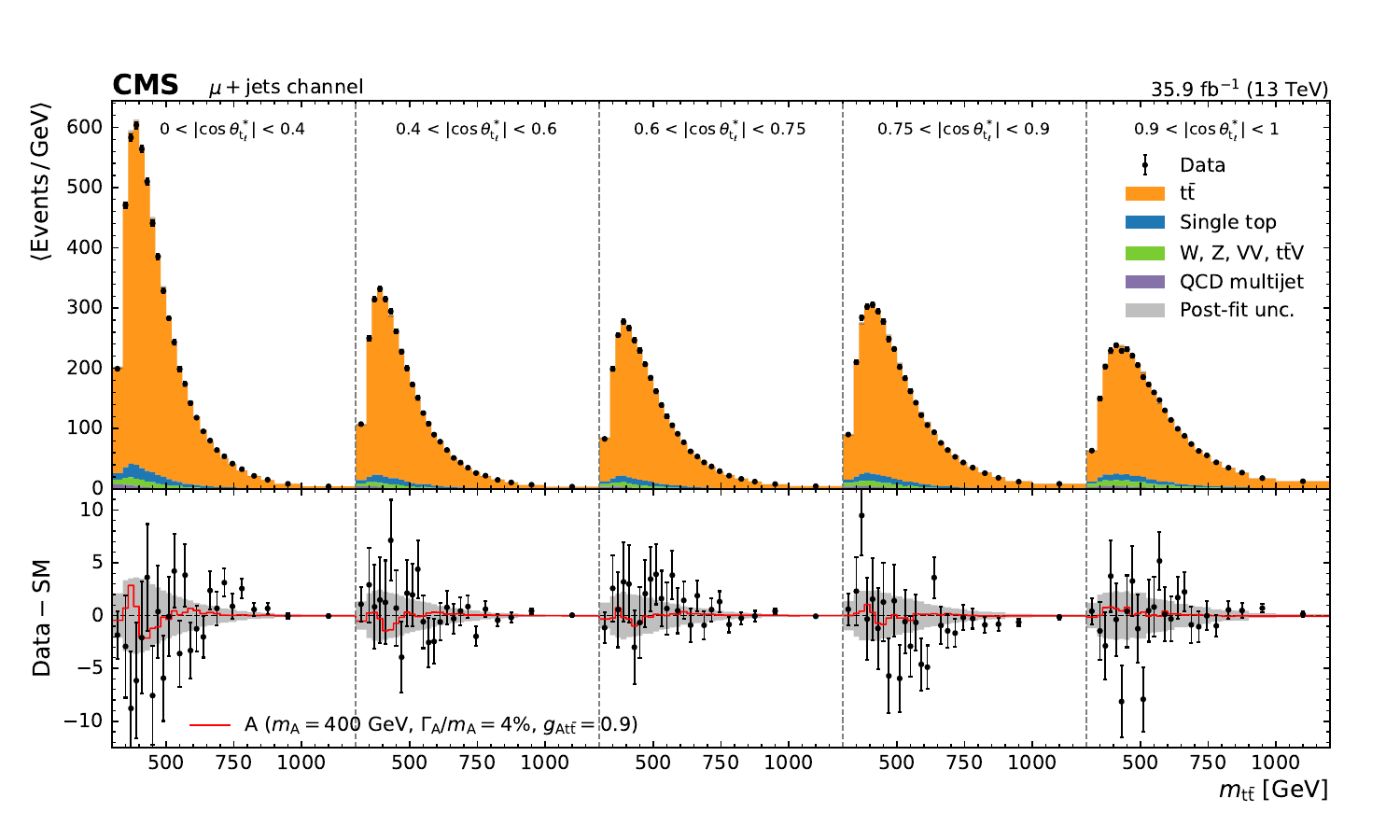}
  \caption{Observed and expected distributions of \mtt{} in different $\abs{\cos\thetaStar}$ regions in the $\Pe + \text{jets}$ (upper) and $\PGm + \text{jets}$ (lower) channels.
  The expected distributions have been obtained with a background-only fit to the data, and an approximate post-fit uncertainty is shown with a gray band.
  The impact of the best-fit signal is included in the lower panels for illustration.}
  \label{fig:SearchVars_singlelep}
\end{figure}

The observed and post-fit predicted distributions of \mtt{} in different $\abs{\cos\thetaStar}$ regions are shown in Fig.~\ref{fig:SearchVars_singlelep}.
The impact of the signal process (including the interference) for the best-fit signal hypothesis, which will be discussed in Section~\ref{sec:modelindependentlimits}, is shown in the lower panels.
It demonstrates the characteristic peak-dip lineshape discussed in Section~\ref{sec:Introduction}.
For this benchmark and also in general, the contributions from both the resonant part and the interference are important.
The relative importance of the latter increases with the total width or as the coupling modifier decreases.

\subsection{Dilepton final state}
\label{sec:dilepanalysis}
In the dilepton channel we aim to select signal events where both top quarks decay to a leptonically decaying \PW~boson.
Hence, the targeted decay topology is $\PQt\PAQt \to \ell^+\PGn \PQb ~ \ell^-\PAGn \PAQb$.
Events in the dielectron ($\Pe\Pe$), electron-muon ($\Pe\Pgm$), and dimuon ($\Pgm\Pgm$) channel in data and simulation are required
to pass a dielectron, electron-muon, and dimuon trigger, respectively, or a single-lepton trigger, as explained in Section~\ref{sec:samples}.
The subsequent event selection closely follows Ref.~\cite{CMS-TOP-17-014}.
Events are required to contain exactly one pair of oppositely charged tight leptons, in which the leading (subleading) lepton has $\pt > 25$ (20)\GeV.
Events are rejected if they contain additional tight electrons or additional muons that satisfy the tight identification criteria with the exception of a looser selection on isolation, $\Irel < 0.25$.
The selected dilepton pair is further required to have an invariant mass of at least 20\GeV, to suppress events from low-mass dilepton resonances.
In the $\Pe\Pe$ and $\mu\mu$ channels, events are rejected if they contain a dilepton pair consistent with a decay of a \PZ~boson,
namely, if they have an invariant mass in the range 76--106\GeV.
Each event must have at least two jets with $\pt > 30$\GeV and $\abs{\eta} < 2.4$.
Additional jets with $\pt > 20$\GeV and $\abs{\eta} < 2.4$ in the event are also considered for further analysis.
At least one of the jets is required to be \cPqb~tagged. In the $\Pe\Pe$ and $\Pgm\Pgm$ channels, the \ptmiss{} must exceed 40\GeV to further suppress $\PZ/\gamma^*$ background events.

The contribution from the $\PZ/\gamma^* + \text{jets}$ background process to the selected event yield is estimated using control regions in data,
following the procedure described in Ref.~\cite{CMS-TOP-11-002}.
The yield outside the \PZ~boson mass window is estimated based on the observed event yield within the window,
using the knowledge from simulation of the ratio of $\PZ/\gamma^* + \text{jets}$ inside and outside of the mass window.
In this estimation, the non-$\PZ/\gamma^* + \text{jets}$ background contribution within the \PZ~boson mass window is taken from the $\Pe\mu$ channel,
where the $\PZ/\gamma^* + \text{jets}$ contribution is negligible, and
corrected for lepton reconstruction effects before being subtracted from the observed number of events in the \PZ~boson mass window.
Using this method, the theoretically predicted $\PZ/\gamma^* + \text{jets}$ event yield is scaled by a factor 1.22, 1.20, and 1.19 in the $\Pe\Pe$, $\Pe\PGm$, and $\PGm\PGm$ channels respectively.

Each event that passes the selection described above is reconstructed under the assumption that it has been produced in a process $\PQt\PAQt \to \ell^+\PGn \PQb ~ \ell^-\PAGn \PAQb$.
A kinematic reconstruction algorithm~\cite{CMS-TOP-12-028} is applied to reconstruct the \ttbar~system. All jets with $\pt > 20$\GeV are considered in the reconstruction of the \ttbar~system.
Given an assignment of jets and leptons to the underlying expected \ttbar~decay products,
a system of equations is constructed that imposes constraints on the reconstructed \PW~boson mass and the reconstructed top quark and antiquark masses.
The transverse momentum imbalance, represented by \ptmiss, is assumed to originate solely from the two neutrinos.
Detector resolution effects are taken into account by sampling both the measured energy and the direction of the leptons and \cPqb~jet candidates within their respective experimental resolutions.
For each sampling, the solution of the equations that results in the smallest \mtt{} is chosen.
Per event, 100 samplings are performed, and each is assigned a weight based on the probability density of the invariant mass of the lepton and \cPqb~jet from the top quark decay.
The kinematic properties of the top quark and antiquark are obtained as a weighted average over all samplings.
Finally, the assignment of jets resulting in the maximum sum of weights over all samplings is chosen, and preference is given to a jet assignment that contains two \cPqb-tagged jets.

The performance of the \ttbar~reconstruction algorithm is studied using simulated SM~\ttbar{} events with targeted decays at the generator level, $\PQt\PAQt \to \ell^+\PGn \PQb ~ \ell^-\PAGn \PAQb$.
In 75\% of the selected events the two generator-level jets are within the acceptance.
For those events for which the algorithm has a solution, the probability to correctly match both \cPqb~jets as chosen by the algorithm
to jets originating from \cPqb~quarks from the top quark decays is 55 to 85\%, depending on the value of $\mtt^\text{gen}$.
The \mtt~resolution, computed using all selected SM~\ttbar events with targeted decays, ranges from 20 to 28\%, depending on $\mtt^\text{gen}$.

The events for which the \ttbar~reconstruction results in a solution, which is the case in about 95\% of observed events, are considered for further analysis.
The resulting event yields for data and SM background expectations are shown in Table~\ref{tab:eventyields_dilep}.

\begin{table}[ht]
\topcaption{
Event yields and composition of SM background in the dilepton channel.
Expected yields are computed in the same way as in Table~\ref{tab:eventyields_singlelep}.
}
\centering
\begin{tabular}{l cc }
\hline
Process  & Event yield  \\
\hline
Observed                      & 230\,233  \\
Total background              & $\left(231.1 \pm 0.8\right)\ten{3}$   \medskip\\
\hline
 & Fraction w.\,r.\,t. total background  \\
\hline
\ttbar                        & 93.3\% \\
Single top quark              & 3.3\%  \\
$\PZ/\gamma^* + \text{jets}$  & 2.9\%  \\
$\ttbar \PV$                  & 0.3\%  \\
Diboson                       & 0.1\% \\
\hline
\end{tabular}
\label{tab:eventyields_dilep}
\end{table}

\begin{figure}
  \centering
  \includegraphics[width=\textwidth]{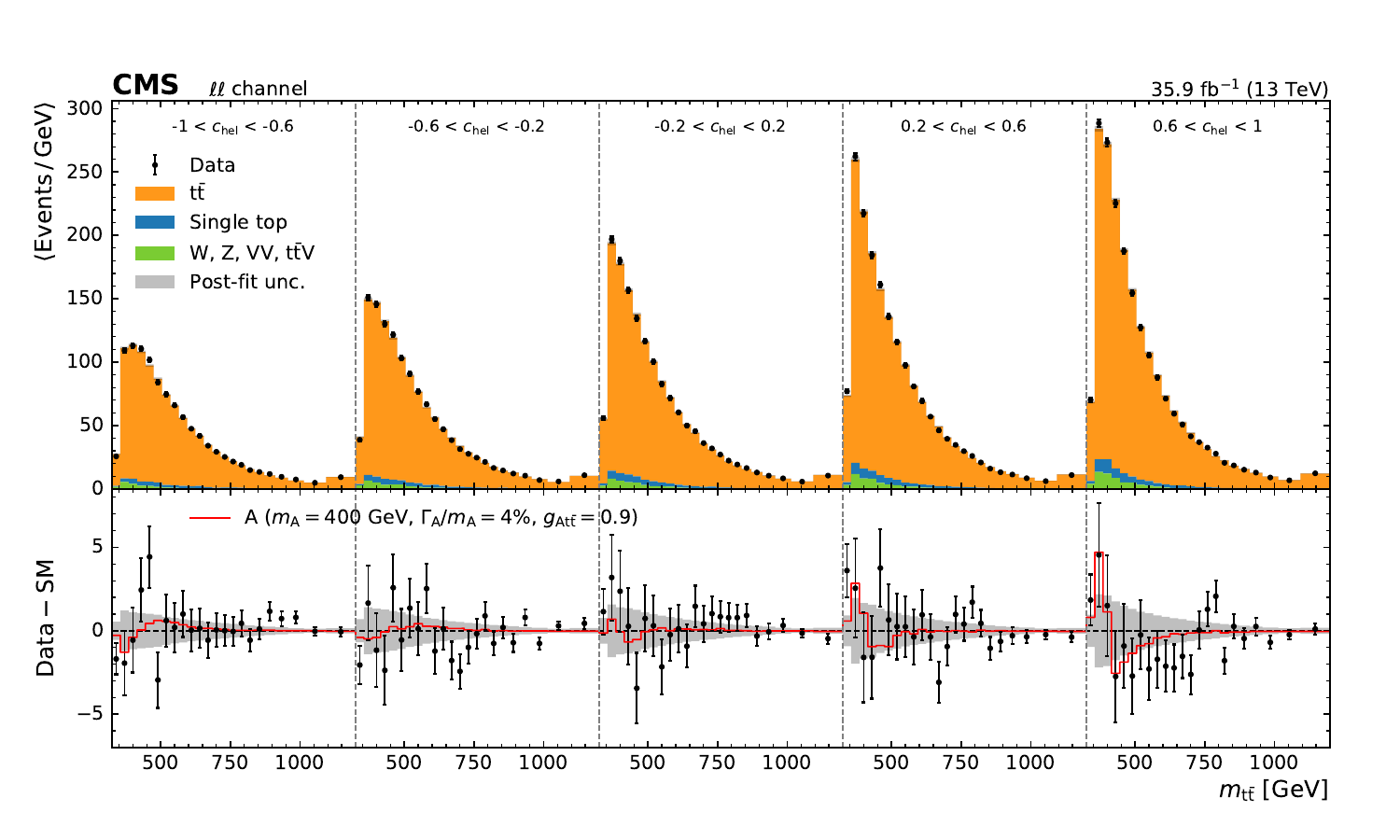}
  \caption{Observed and expected distributions of the observables exploited in the dilepton channel. The expected distributions have been obtained with a background-only fit to the data, and an approximate post-fit uncertainty is shown with a gray band. The impact of the best-fit signal is included in the lower panel for illustration.}
  \label{fig:SearchVars_dilep}
\end{figure}

The search for the $\HH \to \ttbar$ signal is performed using two observables.
The first one is the invariant mass~\mtt, obtained from the \ttbar~reconstruction algorithm.
The second observable is a spin correlation variable constructed from the charged leptons in the event.
Charged leptons have the highest spin analyzing power amongst the top quark decay products~\cite{Bernreuther:2004jv}, and their properties can be measured precisely.
The chosen variable is the cosine of the angle between the charged lepton momenta in their respective helicity frames, and is denoted by $c_\text{hel}$.
The four-momenta of the leptons in their helicity frames are obtained by first boosting the leptons into the \ttbar~rest frame
and then boosting them along their parent top quark or antiquark directions in this frame.
The distribution of $c_\text{hel}$ is sensitive to the spin and CP~state of the \ttbar~system.
At the generator level and with no requirements on acceptance, the distribution is linear in shape; its slope is maximally positive for the \HHOdd~resonance and is mildly negative for the \HHEven~resonance. 
On the other hand, the slope for the SM \ttbar~production, integrated over the entire~\mtt range, is mildly positive.
This allows discriminating both between the signal and background processes and between the \HHEven{} and \HHOdd~states.
The observed and post-fit predicted distributions used in the dilepton channel are shown in Fig.~\ref{fig:SearchVars_dilep}.

\section{Systematic uncertainties}
\label{sec:systematics}
Various sources of uncertainty affect the distributions of the observables used to search for a heavy Higgs boson signal.
Below we describe the experimental and theoretical systematic effects considered in the analysis.
In the statistical evaluation discussed in Section~\ref{sec:Results},
each source of uncertainty corresponds to a nuisance parameter in a binned maximum-likelihood fit to the distributions of the observables in data.
Uncertainties that affect only the normalization are modeled using log-normal constraints, while Gaussian constraints are imposed for nuisance parameters that control all other uncertainties.
Unless stated otherwise, all uncertainties are evaluated on signal as well as background processes and treated as fully correlated among the processes and
lepton channels.
The uncertainties are summarized in Table~\ref{tab:systematics}.

The uncertainty due to the jet \pt~scale~\cite{Khachatryan:2016kdb} is evaluated by varying the corresponding corrections within their uncertainties.
The events are reanalyzed, by reapplying the event selection and recalculating all kinematic quantities.
A total of 19 independent jet momentum correction uncertainties affecting jets in the tracker acceptance are considered.
As the jet \pt{} resolution in simulation is smeared to match the resolution observed in data, a corresponding uncertainty is evaluated.
An uncertainty in the unclustered component of \ptmiss{} is computed by shifting the energies of PF candidates not clustered into jets with $\pt > 15$\GeV
according to the energy resolution for each type of PF candidate~\cite{Sirunyan:2019kia}.
Uncertainties in the \cPqb~tagging efficiency scale factors applied to simulated events are evaluated by varying them within the respective uncertainties~\cite{CMS-BTV-16-002}.
The scale factors for heavy-flavor (\cPqb{} and \cPqc) jets are varied independently of those for light-flavor jets.
The uncertainties in the trigger scale factors as well as the electron and muon identification scale factors are considered~\cite{Khachatryan:2015hwa,Sirunyan:2018fpa},
where the lepton identification also includes effects originating from the isolation requirement and the track reconstruction.
The uncertainties in the trigger efficiency scale factors for the single-electron and single-muon channels are considered not correlated with each other, but each of them is independently correlated with the uncertainty in the trigger scale factors in the dilepton channel, with a 50\% correlation coefficient.
Effects due to the uncertainty in the distribution of the number of pileup interactions
are evaluated by varying the effective inelastic proton-proton cross section
in the simulation by 4.6\% from its nominal value.
The uncertainty in the integrated luminosity amounts to 2.5\%~\cite{CMS-PAS-LUM-17-001} and affects the normalization of all simulated processes.

The prediction of the SM \ttbar{} production, the main background process in the analysis, is affected by various sources of theoretical uncertainties.
The overall normalization of the SM \ttbar{} background is assigned an uncertainty of 6\%, from the $\text{NNLO} + \text{NNLL}$ QCD cross section calculation~\cite{Czakon:2011xx,Botje:2011sn} that is used to normalize the events.
The effect of the choice of the renormalization and factorization scales, $\mu_\text{R}$ and $\mu_\text{F}$, in the matrix element
is evaluated by varying these scales independently by a factor of 2 and $1/2$.
The effect on the acceptance is considered, but the effect on the cross section is ignored as it is already included in the considered uncertainty in the cross section.
The renormalization scales used in the parton shower simulation of initial-state radiation (ISR) and final-state radiation (FSR) are also varied independently by a factor of 2 in each direction.
The effect of the uncertainty in the amount of ISR, as well as in the underlying event tune used in the simulation, was found not to be statistically significant.
The uncertainty in the top quark mass is considered by shifting $m_{\cPqt}$ in the simulation by $\pm 3$\GeV
and rescaling the induced variations by a factor of $1/6$ to emulate a more realistic top quark mass uncertainty of 0.5\GeV~\cite{Khachatryan:2015hba}.
The uncertainty in the matching scale between the matrix element and the parton shower is evaluated
by varying the \POWHEG parameter, $h_{\text{damp}}$, that controls the suppression of radiation of additional high-\pt jets~\cite{CMS-PAS-TOP-16-021}.
The nominal value of $h_{\text{damp}}$ in the simulation is $1.58~m_{\cPqt}$, and the varied values are $0.99~m_{\cPqt}$ and $2.24~m_{\cPqt}$.
The uncertainty arising from the choice of the PDF set is evaluated by reweighting the simulated \ttbar~events using 100 replicas of the NNPDF3.0 set.
A principal component analysis is performed on the variations from the PDF replicas to construct two base variations,
such that the deviation from the nominal distribution given by each replica can be described as a linear combination of the base variations.
The uncertainty in the \alpS~parameter used in the PDF set induces a third independent PDF variation.
The uncertainty accounting for the mismodeling of the \pt~spectrum of top quarks is evaluated
by varying the two parameters used in the top quark \pt{} reweighting function.

The renormalization and factorization scale uncertainties in the heavy Higgs boson signal simulation
are treated independently for the resonant and interference components.
Compared to the alternative of varying the scales for the two components simultaneously, we found this to be the more conservative option.
The effect on the acceptance as well as on the cross section is considered.
Since the simulated samples have been generated at LO accuracy, the total effect on the cross section reaches values in excess of 30\%.
Other theoretical uncertainties in the signal, such as the uncertainties in $m_{\cPqt}$ or PDF, are neglected as they are expected to be small compared to this variation.

The expected yields for most of the non-\ttbar{} background processes are derived using theoretical predictions for the cross sections at NLO or higher accuracy.
The uncertainties assumed in the normalization of these processes are conservative and always exceed those of the corresponding theoretical computations.
For the single top quark production in the $t$ ($\cPqt\PW$) channel we assign an uncertainty of 20 (15)\%, which is based on the measurements of the cross sections of these processes~\cite{Aaboud:2016ymp,Sirunyan:2018rlu,Sirunyan:2018lcp}.
The uncertainty in the $\ttbar\PV$ production is taken to be 30\%, which covers the uncertainties of the experimental measurements~\cite{Sirunyan:2017uzs,Aaboud:2019njj}.
To account for the fact that this search probes a restricted region of the phase space of the corresponding processes, we assign uncertainties of 50\% for $\PW + \text{jets}$ and $\PZ/\gamma^* + \text{jets}$ production (only in the single-lepton channel) and 30\% for the diboson production.
Finally, a 20\% uncertainty is used for the $s$-channel single top quark production, for which no measurement at the LHC exists.
The adopted conservative normalization uncertainties have little impact on the sensitivity of this search due to the small contribution of these processes.

In cases where the normalization of a background process is estimated using a data-driven method, the corresponding uncertainty is determined by the same method.
For the Drell--Yan background in the dilepton channel we assign a 30\% uncertainty, from the variation in the scale factors when derived with and without the requirements on \ptmiss{}, the jet \cPqb~tag decisions, and the \ttbar~reconstruction.
The normalization of the QCD multijet background, which is only relevant in the single-lepton channel, is assigned an uncertainty of ${+100} / {-50}$\%, independently in the single-electron and single-muon channels.
It covers the statistical uncertainty in the underlying fit and the difference between the data-driven and MC-based estimations.

The nominal background prediction is affected by the limited size of the simulated MC event samples.
This statistical uncertainty is evaluated using the ``light'' Barlow--Beeston method~\cite{barlowbeeston}, by introducing one additional nuisance parameter per bin of the 2D distribution of the observables.

\begin{table}[hbtp]
  \centering
  \topcaption{The systematic uncertainties considered in the analysis, indicating the number of corresponding nuisance parameters (when more than one) in the statistical model,
  the type (affecting shape or only normalization), the affected processes, and the correlation among the lepton channels.
  Uncertainties tagged in the last column with ``All'' are correlated among the single-lepton and dilepton channels.
  In case an uncertainty is only applicable to the single-electron, the single-muon, the single-lepton, or the dilepton channel,
  they are indicated with \Pe, \Pgm, $\ell$, $\ell\ell$, respectively.
  }
  \label{tab:systematics}
  \renewcommand{\arraystretch}{1.2}  \begin{tabular}{lllc}
  \hline
  \multicolumn{1}{l}{Uncertainty (\# of parameters)}    & \multicolumn{1}{l}{Type} & \multicolumn{1}{l}{Affected process} & \multicolumn{1}{l}{Correlation}   \\
  \hline
   Jet \pt{} scale (19)               &   shape          &    All    &  All    \\
   Jet \pt{} resolution               &   shape          &    All    &  All    \\
   Unclustered \ptmiss                &   shape          &    All    &  All    \\
   \cPqb tagging heavy-flavor jets    &   shape       &  All     &   All    \\				
   \cPqb tagging light-flavor jets    &   shape       &  All     &   All     \\	    	
   Pileup                             &   shape       &  All     &   All     \\					
   Electron identification            &   shape      &   All     &   All      \\
   Muon identification                &   shape       &   All    &   All      \\
   Single-electron trigger                &   shape        &    All      &  \Pe, $\ell\ell$    \\
   Single-muon trigger                    &   shape       &   All        &  \Pgm, $\ell\ell$    \\	
   Luminosity calibration                 &   norm.   &  All     &  All     \\[\cmsTabSkip]
   Renorm. scale SM \ttbar                   &   shape       &     SM \ttbar     &    All    \\
   Fact. scale SM \ttbar                     &   shape       &     SM \ttbar     &    All    \\   	    	
   Parton shower FSR \ttbar                  &   shape       &     SM \ttbar     &    All    \\					
   $h_{\text{damp}}$                         &   shape       &     SM \ttbar     &    All    \\
   Top quark mass                            &   shape       &     SM \ttbar     &    All    \\
   Top quark \pt{} (2)                           &   shape       &     SM \ttbar     &    All    \\            				
   PDF (3)                                       &   shape       &     SM \ttbar     &    All   \\
   Renorm. scale res. signal                 &   shape       &     Resonant signal        &    All    \\
   Renorm. scale int. signal                 &   shape       &     Interference signal    &    All    \\
   Fact. scale res. signal                   &   shape       &     Resonant signal        &    All    \\
   Fact. scale int. signal                   &   shape       &     Interference signal    &    All    \\[\cmsTabSkip]
   SM \ttbar norm.                     &    norm.      &   SM \ttbar    &    All    \\
   Single top $t$ channel norm.                   &    norm.      &  Single top $t$ channel     &    $\ell$    \\
   Single top $s$ channel norm.                   &    norm.      &  Single top $s$ channel     &    $\ell$    \\
   Single top $\cPqt\PW$ channel norm.            &    norm.      &  Single top tW channel        &     All     \\
   $\PW + \text{jets}$ norm.                      &    norm.      &   $\PW + \text{jets}$         &     $\ell$    \\
   $\PZ/\gamma^* + \text{jets}$ norm.                     &    norm.      &   $\PZ/\gamma^* + \text{jets}$         &    $\ell$    \\
   $\PZ/\gamma^* + \text{jets}$ norm. from data                     &    norm.      &   $\PZ/\gamma^* + \text{jets}$    &    $\ell\ell$    \\
   Diboson norm.                      &    norm.      &   Diboson         &    All    \\                	
   $\ttbar \PV$ norm.                 &    norm.      &   $\ttbar \PV$    &    All    \\       	
   QCD multijet norm. from data, \Pe       &    norm.      &   QCD multijet    &  \Pe    \\
   QCD multijet norm. from data, \Pgm      &    norm.      &   QCD multijet    &  \Pgm    \\[\cmsTabSkip]
   MC statistical uncertainty (365)          &   shape       &     All     &    No    \\     				
  \hline
  \end{tabular}
\end{table}

Several systematic variations in the background, most notably those constructed from dedicated MC samples, are affected by statistical fluctuations.
We suppress these fluctuations by smoothing the relative deviations from the nominal background distribution of \mtt{} and the angular variable.
The up and down deviations for each independent uncertainty are assumed to be symmetric in shape, but allowed to differ in the overall size.
The symmetrized deviation is smoothed using a version of the LOWESS algorithm (LOcally WEighted Scatterplot Smoothing)~\cite{Cleveland79, Cleveland88}.
In the vicinity of each bin of the 2D distribution, the symmetrized deviation is fitted with a plane using a weighted least squares fit, in which nearby bins receive larger weights.
The smoothed deviation obtained in this way is rescaled to account for the overall size of the input up or down deviation, and applied to the nominal background expectation in the given bin.
A similar procedure is also applied to all signal distributions.
The resulting distributions are then used in the subsequent analysis to evaluate the systematic uncertainty under consideration.

In general, the relative importance of different systematic uncertainties depends greatly on the signal hypothesis, especially the mass of the heavy Higgs boson.
Typically, among the uncertainties with the largest impact are the signal theoretical uncertainties and some of the jet momentum correction uncertainties.
Close to the \ttbar{} production threshold, the variations in $m_{\cPqt}$ and the $h_\text{damp}$ parameter become important, while for larger $m_\HH$ the PDF, $\mu_\text{R}$, and $\mu_\text{F}$ variations in the SM~\ttbar{} background can have significant impacts.
For certain signal hypotheses, the signal distribution partially resembles one of a minor background; in such cases the variation of the normalization of the respective background becomes relevant.
In addition, MC statistical uncertainties, when grouped together, often outweigh every other individual uncertainty.

\section{Results}
\label{sec:Results}
To evaluate the consistency of the observed data with the presence of a signal, we perform a statistical analysis using the 2D binned distribution of $(\mtt, \abs{\cos\thetaStar})$
in the single-electron and the single-muon channels separately, and the 2D binned distribution of $(\mtt, c_\text{hel})$ in the combined dilepton channel.
The single-lepton and dilepton channels do not overlap as they correspond to orthogonal lepton selection criteria.

The statistical model is defined by the likelihood function
\begin{linenomath}
\begin{equation}
 \label{Eq:likelihood}
 \begin{gathered}
 L(\mu, \mathbf{p}, \boldsymbol{\nu}) = \left(\prod_i \frac{\lambda_i^{n_i}(\mu, \mathbf{p}, \boldsymbol{\nu})}{n_i!}\, \re^{-\lambda_i(\mu, \mathbf{p}, \boldsymbol{\nu})}\right) \, G(\boldsymbol{\nu}),\\
 \lambda_i(\mu, \mathbf{p}, \boldsymbol{\nu}) = \mu \sum_{\HH = \HHEven, \HHOdd} \left(g_{\HH\ttbar}^4\,
 s_{R,i}^\HH(m_\HH, \Gamma_\HH, \boldsymbol{\nu}) + g_{\HH\ttbar}^2\, s_{I,i}^\HH(m_\HH, \Gamma_\HH, \boldsymbol{\nu})\right) + b_i(\boldsymbol{\nu}),
 \end{gathered}
\end{equation}
\end{linenomath}
with $b_i$ denoting the combined background yield in a given bin $i$, $s_{R,i}^\HH$ and $s_{I,i}^\HH$ the signal yields in a given bin for the resonant and interference part, respectively,
$\boldsymbol{\nu}$ the vector of nuisance parameters (on which the signal and background yields generally depend), $n_i$ the observed yield, and $g_{\HH\ttbar}$ the coupling strength modifiers given by Eq.~\eqref{Eq:coupling}.
The parameters of the signal model (mass~$m_\HH$, width~$\Gamma_\HH$, and $g_{\HH\ttbar}$) are collectively denoted by vector~$\mathbf{p}$.
Eq.~\eqref{Eq:likelihood} is kept generic by including contributions from both CP states.
As there is no interference between them, the corresponding signal distributions are trivially added together.
We also introduce an auxiliary overall signal strength modifier~$\mu$, which rescales the full beyond the SM (BSM) contribution.
This allows testing different signal hypotheses in a computationally efficient way, as will be detailed below.
The external constraints on the nuisance parameters are taken into account in this likelihood
via a product of corresponding probability density functions, $G(\boldsymbol{\nu})$.

The background-only model is constructed by setting $\mu = 0$ in Eq.~\eqref{Eq:likelihood}.
To quantify the level of agreement between it and observed data, we perform a goodness-of-fit test based on the so-called ``saturated model''~\cite{Baker:1983tu}.
This yields a $p$-value of 0.43, indicating a good overall agreement.

We perform scans over the parameters of the signal models, $\mathbf{p}$.
A variant of the LHC profile likelihood ratio test statistic~$\tilde q_\mu$ from Refs.~\cite{Cowan:2010js,CMS-NOTE-2011-005} is utilized:
\begin{linenomath}
\begin{equation}
 \label{Eq:TestStat}
 \tilde q_{\mu,\mathbf{p}} = -2 \ln \frac{L(\mu, \mathbf{p}, \hat{\boldsymbol{\nu}}_{\mu,\mathbf{p}})}{L(\hat\mu_\mathbf{p}, \mathbf{p}, \hat{\boldsymbol{\nu}}_\mathbf{p})}, \quad 0 \leqslant \hat\mu_\mathbf{p} \leqslant \mu.
\end{equation}
\end{linenomath}
The test statistic is expressed in terms of the auxiliary parameter~$\mu$ in Eq.~(\ref{Eq:likelihood}), in which the statistical model is linear, while the parameters~$\mathbf{p}$ are kept fixed at their values being probed in the scan.
The likelihood in the numerator is maximized with respect to the nuisance parameters, and $\hat{\boldsymbol{\nu}}_{\mu,\mathbf{p}}$ denotes the vector of their values at the maximum.
A similar notation is used in the denominator, where the likelihood is maximized with respect to both $\mu$ and $\boldsymbol{\nu}$, under the additional constraint $0 \leqslant \hat\mu_\mathbf{p} \leqslant \mu$.
The requirement $\hat\mu_\mathbf{p} \geqslant 0$ excludes from the consideration cases in which the shape of the overall BSM contribution gets flipped, resulting in a qualitatively different effect from what is targeted in this search.
The condition $\hat\mu_\mathbf{p} \leqslant \mu$ prevents the exclusion of a signal hypothesis if the data are more compatible with a model that predicts the BSM contribution of a similar shape but a larger overall size.

For each signal hypothesis~$\mathbf{p}$, we perform a test according to the \CLs{} criterion~\cite{Junk:1999kv,Read:2002hq}.
This is done for $\mu = 1$ in Eq.~\eqref{Eq:TestStat}, which reproduces the nominal signal expectation.
We profit from the known asymptotic approximation~\cite{Cowan:2010js} for distributions of the adopted test statistic to construct these distributions in a computationally efficient way.
If the \CLs{} value computed for $\mu = 1$ and given $\mathbf{p}$ is found to be smaller than 0.05, the point~$\mathbf{p}$ is said to be excluded at 95\% confidence level (\CL).

\subsection{Interpolation and extrapolation of signal masses and widths}
\label{sec:morphing}
To construct expected signal distributions for every point encountered in the scans, we apply an interpolation in mass and width of the heavy Higgs boson, starting from the reference generated points.
This is done independently for the \mtt~distribution in each bin of the angular variable.
We consider the resonant part of the signal and the interference separately and further split the interference contribution in two according to the sign of the per-event generator weight.
A change in the mass results in a horizontal shift of the \mtt~distribution.
The interpolation in this observable is implemented with a nonlinear morphing algorithm~\cite{RooMomentMorph}.
On the other hand, the effect of a change in $\Gamma_\HH$ is evaluated with an independent interpolation in each bin.

The signal model parameter scan may reach values of $\Gamma_\HH / m_\HH$ below 2.5\%, the lowest value considered in the simulated signal samples.
Since the reconstructed \mtt~resolution is about 17\% or worse,
the shape of the \mtt~distribution for a signal with such low widths does not differ from the one corresponding to $\Gamma_\HH / m_\HH = 2.5\%$.
Hence, for scan points with $\Gamma_\HH / m_\HH < 2.5\%$ it is sufficient to use the distributions for $\Gamma_\HH / m_\HH = 2.5\%$
and only scale the cross sections appropriately.

\subsection{Model-independent interpretation}
\label{sec:modelindependentlimits}
Constraints on the coupling strength modifier $g_{\HH\ttbar}$ are derived as a function of the mass and width of the heavy Higgs boson, for each CP state independently.
The coupling modifier for the other CP state in Eq.~\eqref{Eq:likelihood} is set to zero to exclude it from the statistical model.
The scan is performed for $m_\HH$ between 400 and 750\GeV and $\Gamma_\HH / m_\HH$ between 0.5 and 25\%.
The mass and width interpolation described in Section~\ref{sec:morphing} is performed in scan points other than those corresponding to the generated signal samples.
Coupling strength values up to 3 are probed to guarantee that the amplitudes preserve perturbative unitarity for all calculations, in accordance with the lower bound $\tan\beta = 1 / g_{\HHOdd\ttbar} \gtrsim 0.3$ given in Ref.~\cite{Branco:2011iw} in the context of 2HDMs.

\begin{figure}[p]
\centering
\includegraphics[width=0.39\textwidth,keepaspectratio=true]{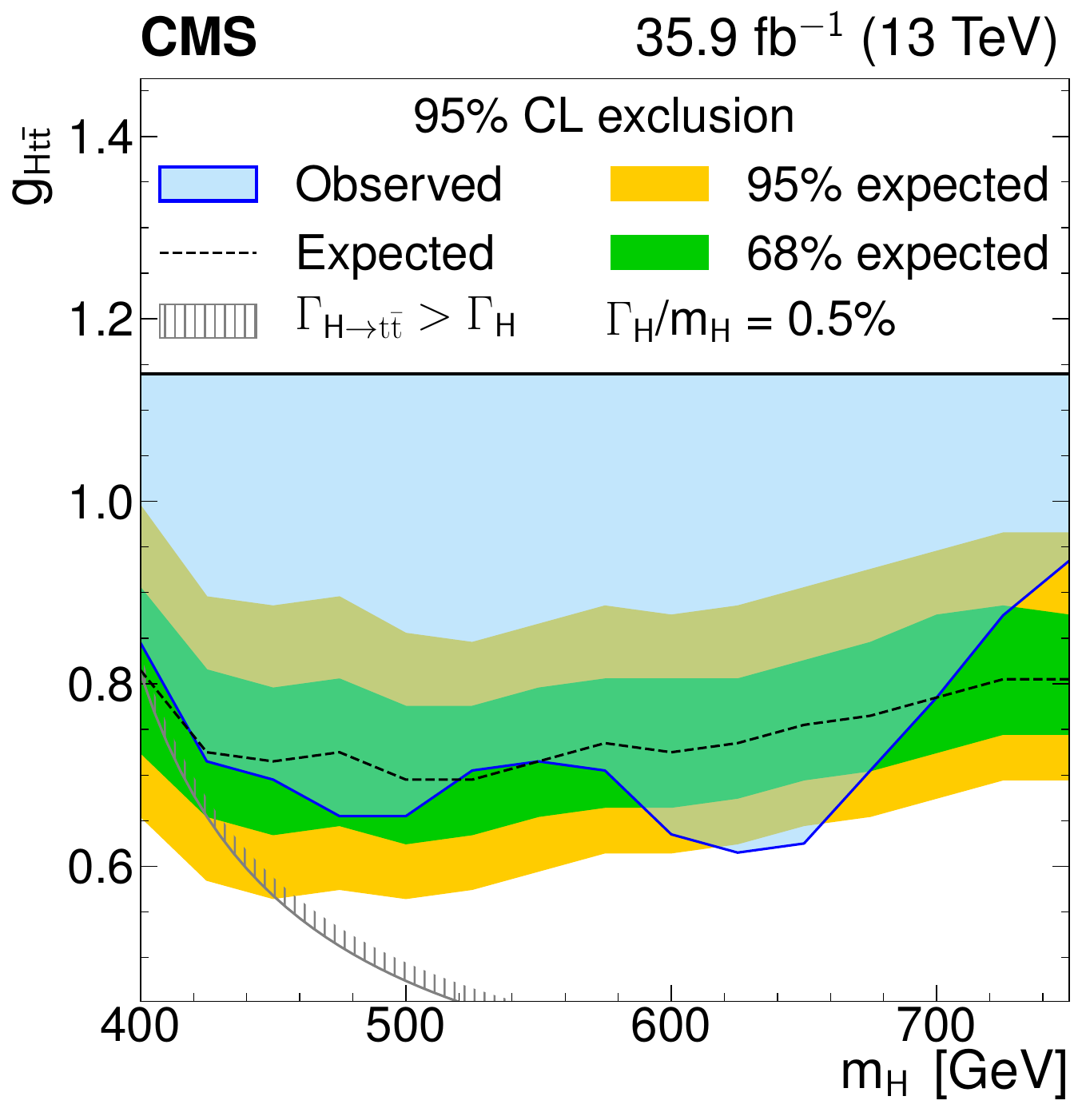}
\includegraphics[width=0.39\textwidth,keepaspectratio=true]{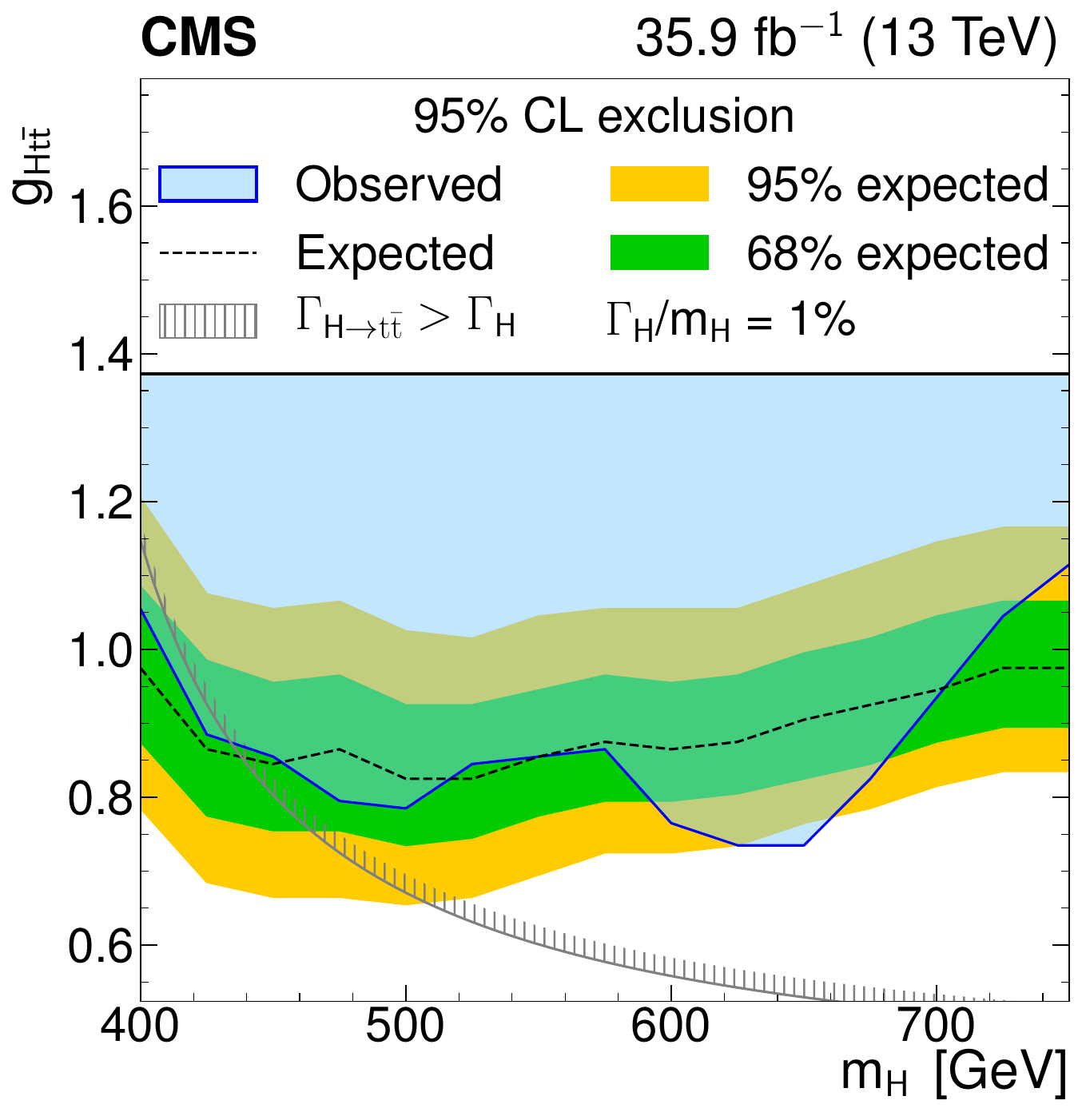}
\includegraphics[width=0.39\textwidth,keepaspectratio=true]{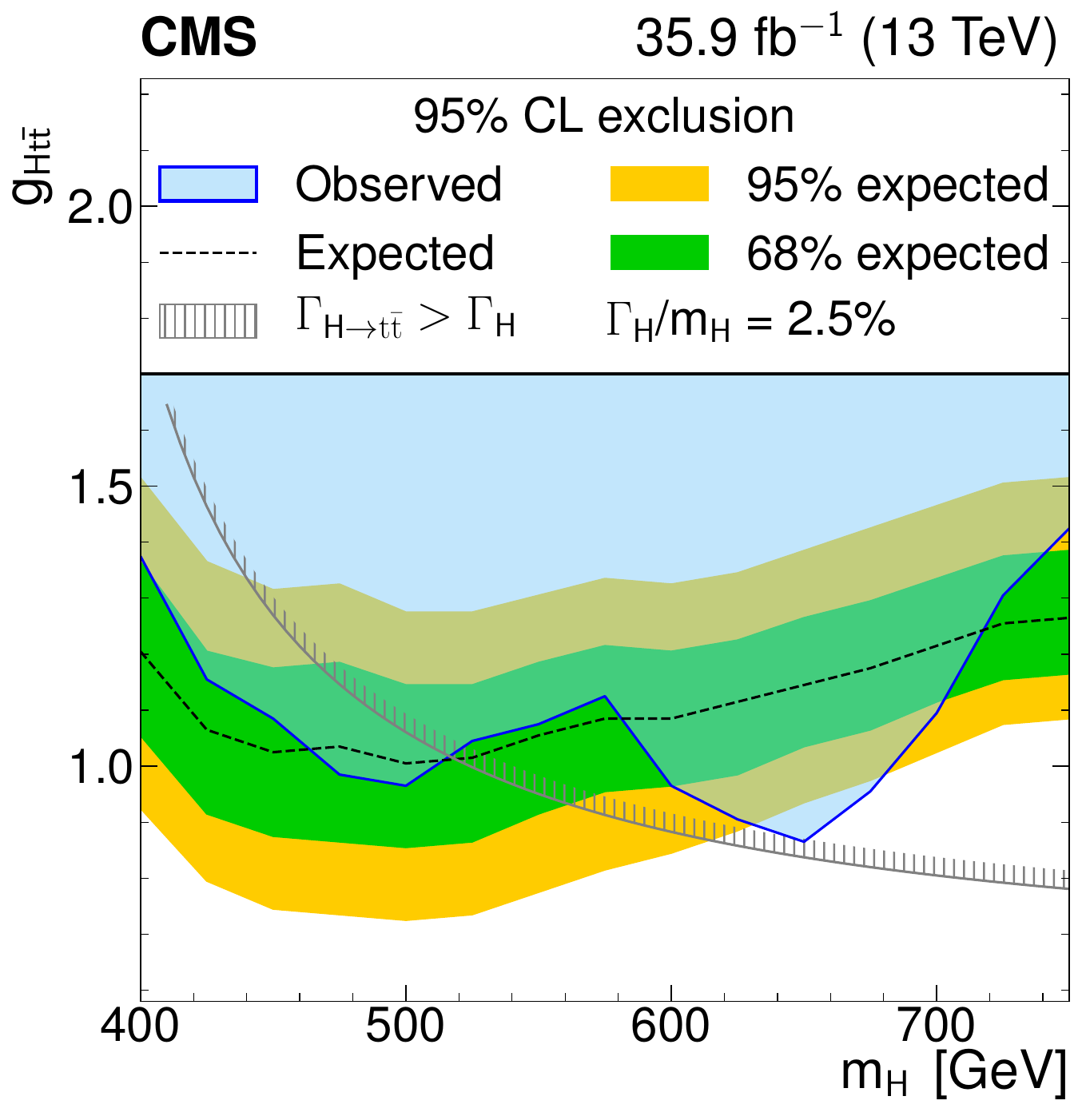}
\includegraphics[width=0.39\textwidth,keepaspectratio=true]{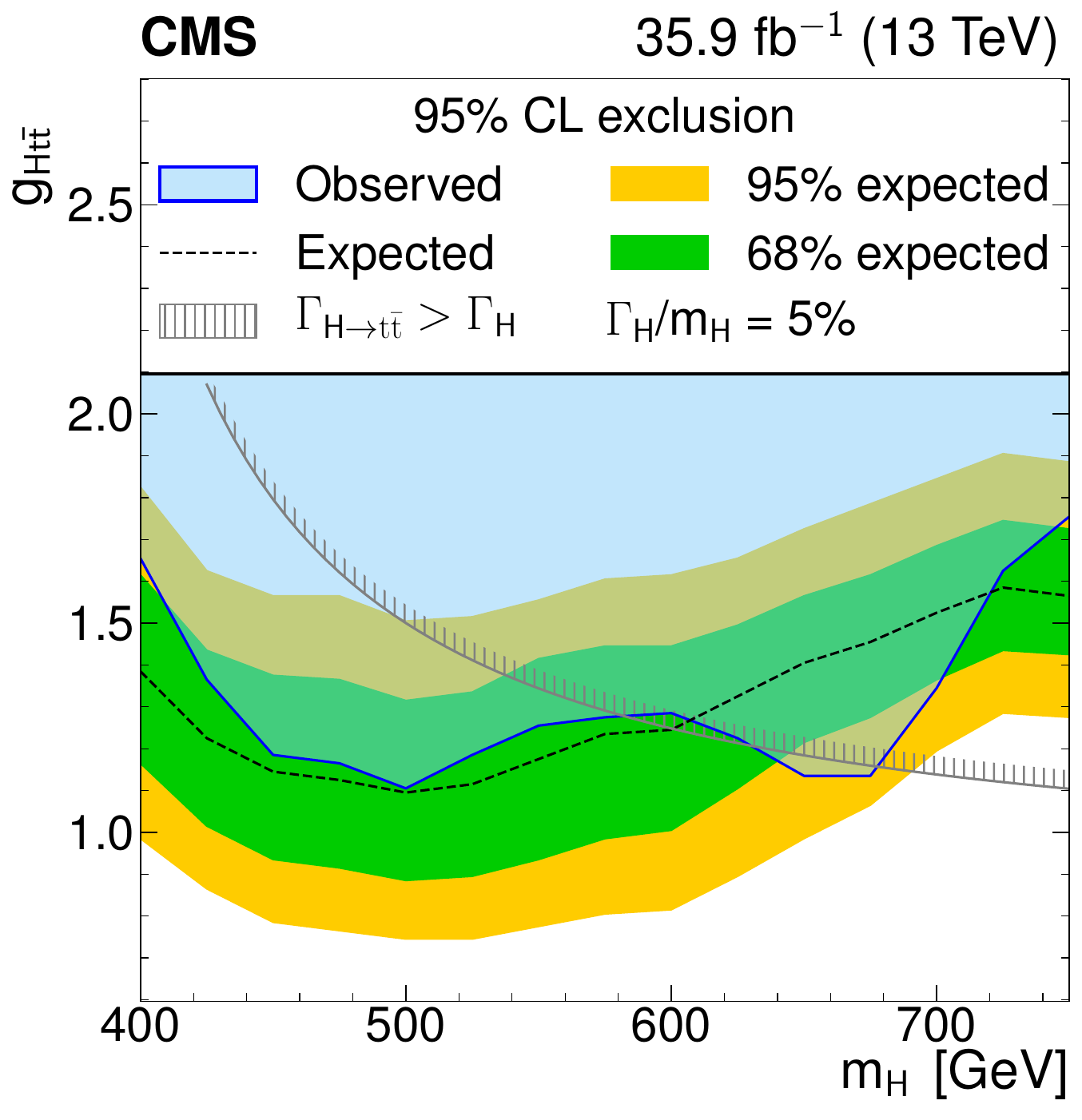}
\includegraphics[width=0.39\textwidth,keepaspectratio=true]{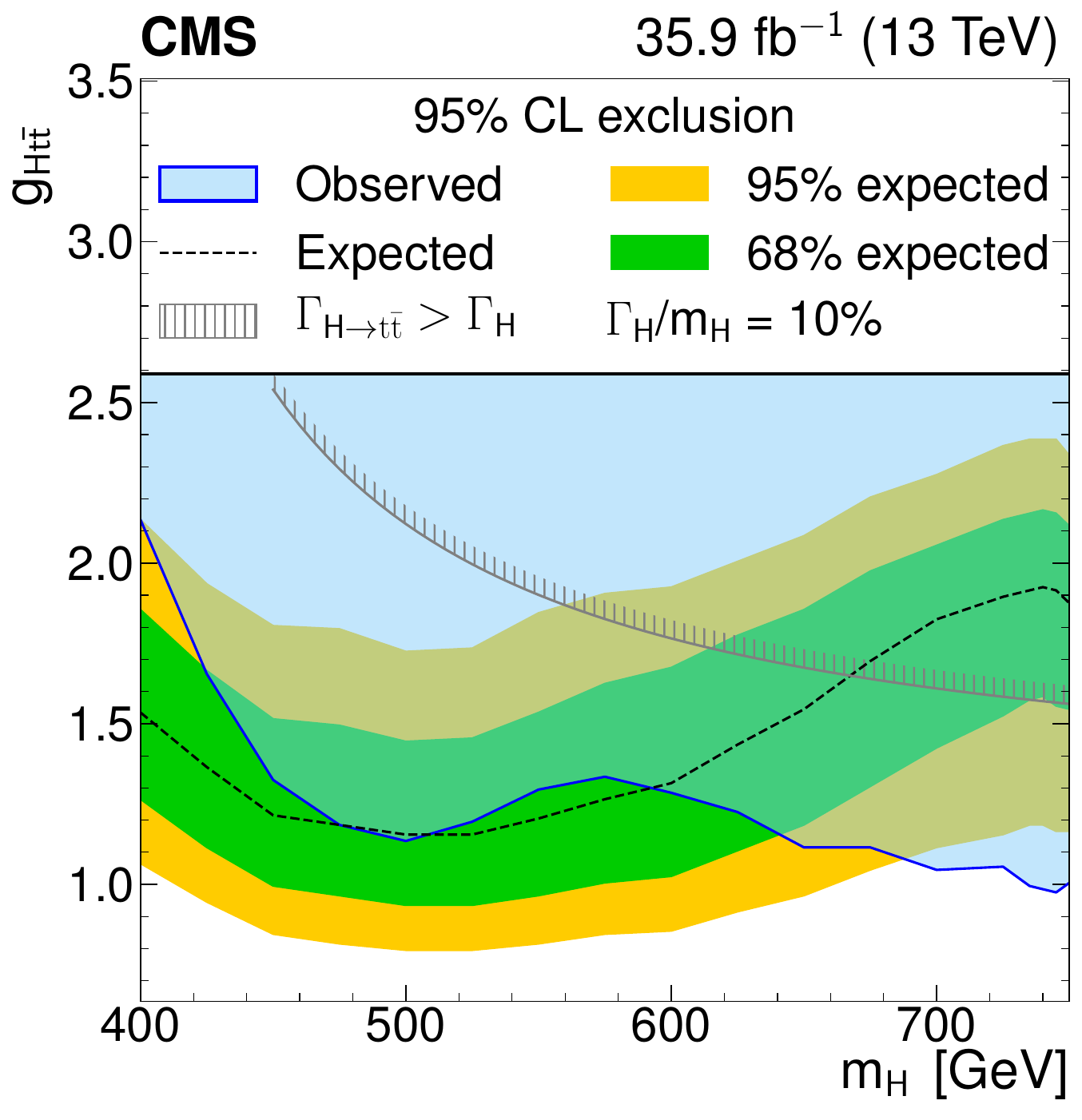}
\includegraphics[width=0.39\textwidth,keepaspectratio=true]{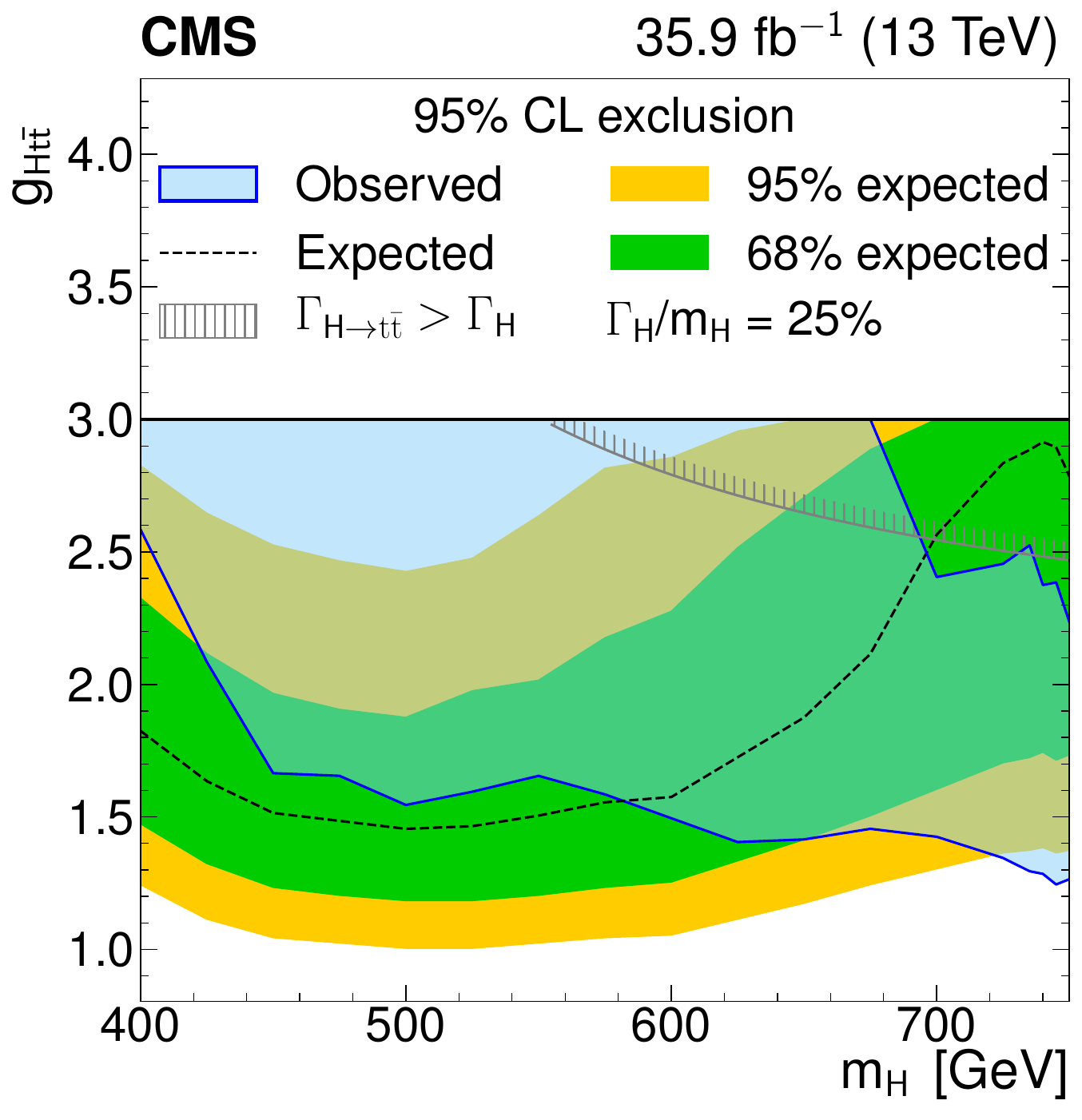}
\caption{Model-independent constraints on the coupling strength modifier as a function of the heavy scalar boson mass, for relative widths of 0.5, 1, 2.5, 5, 10, and 25\%.
The observed constraints are indicated by the blue shaded area.
The inner (green) band and the outer (yellow) band indicate the regions containing 68 and 95\%, respectively, of the distribution of constraints expected under the background-only hypothesis.
The unphysical region of phase space in which the partial width $\Gamma_{\HHEven \to \ttbar}$ becomes larger than the total width is indicated by the hatched lines.}
\label{fig:limits_h_widths}
\end{figure}

\begin{figure}[p]
\centering
\includegraphics[width=0.39\textwidth,keepaspectratio=true]{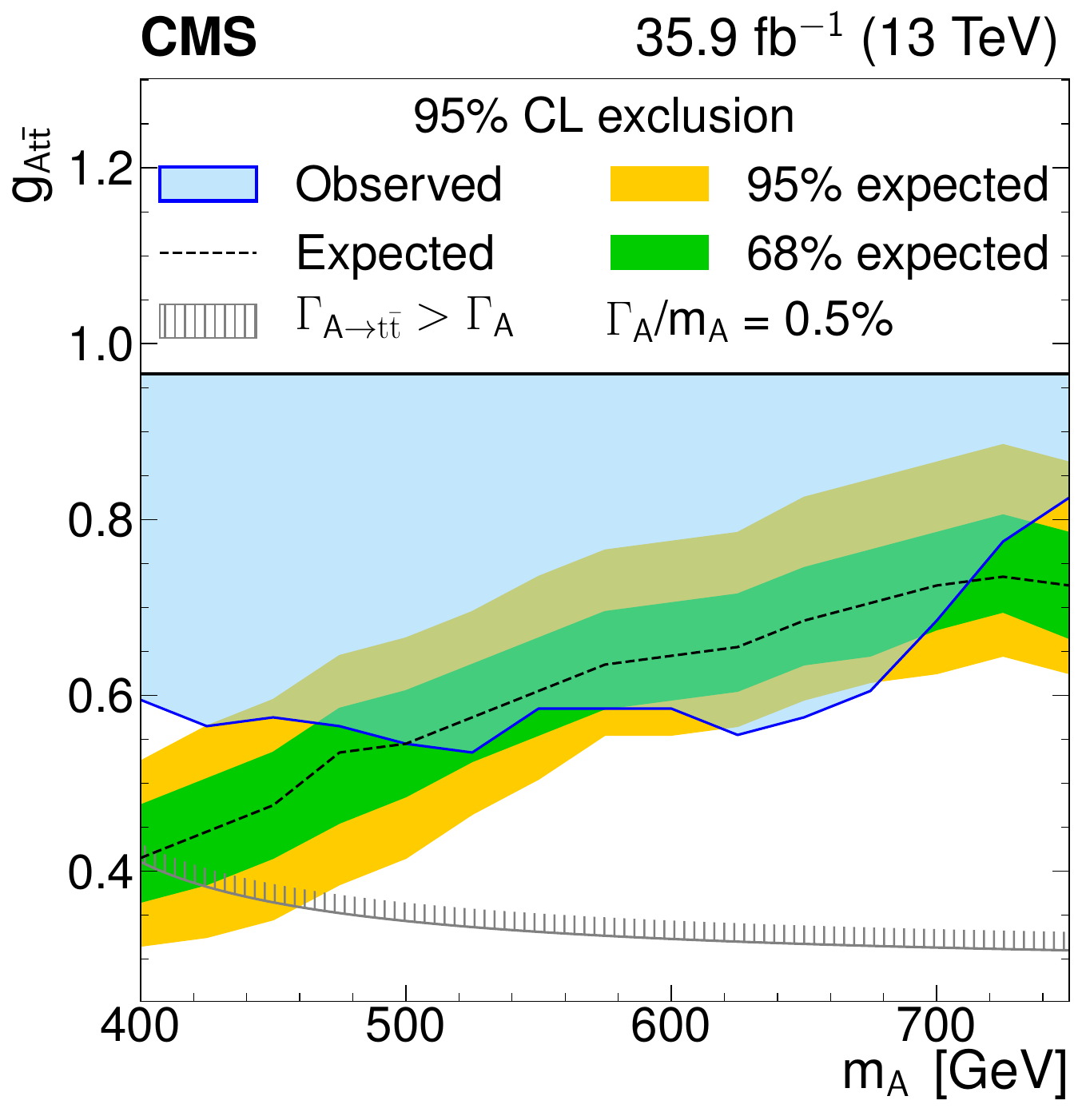}
\includegraphics[width=0.39\textwidth,keepaspectratio=true]{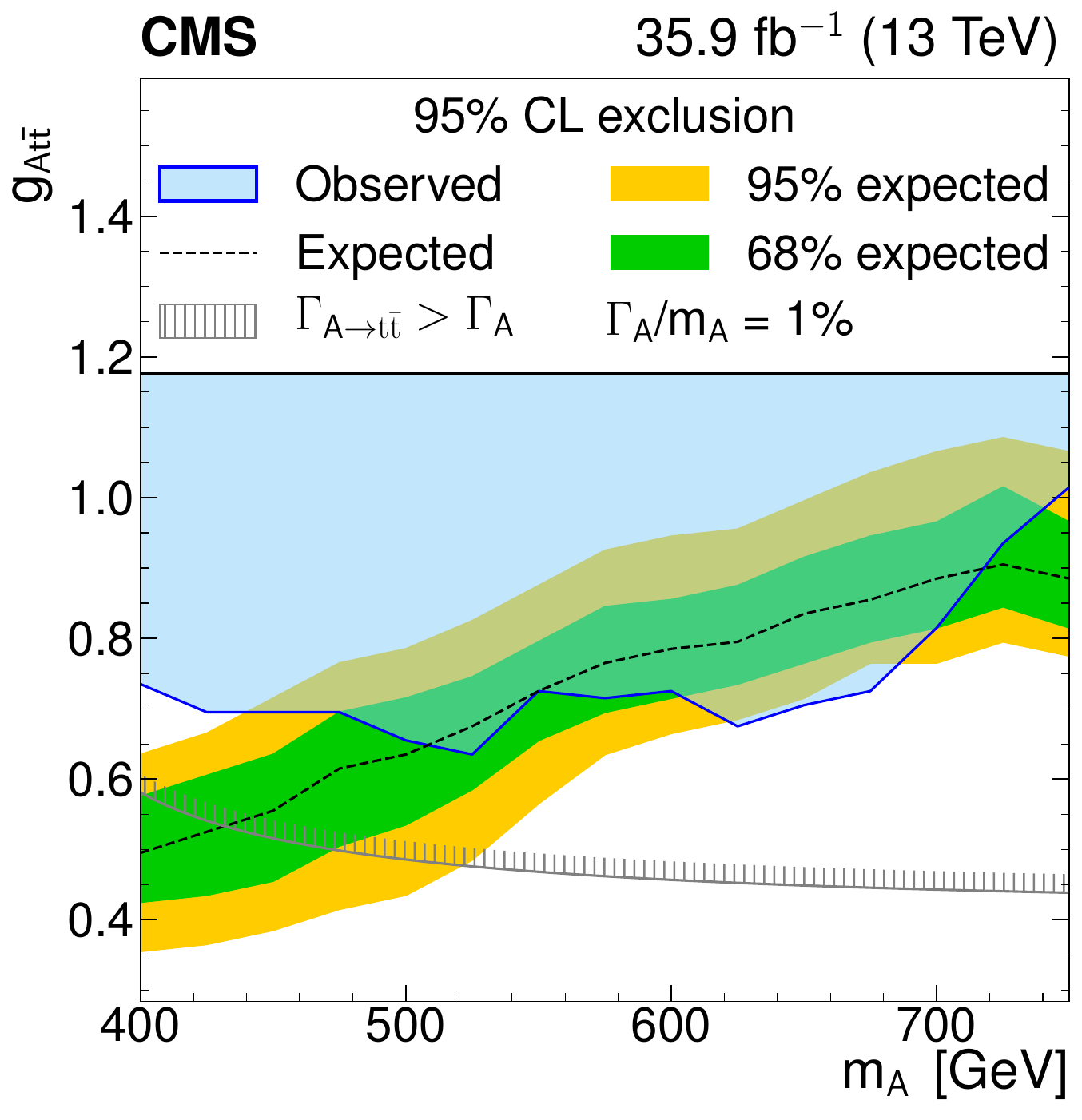}
\includegraphics[width=0.39\textwidth,keepaspectratio=true]{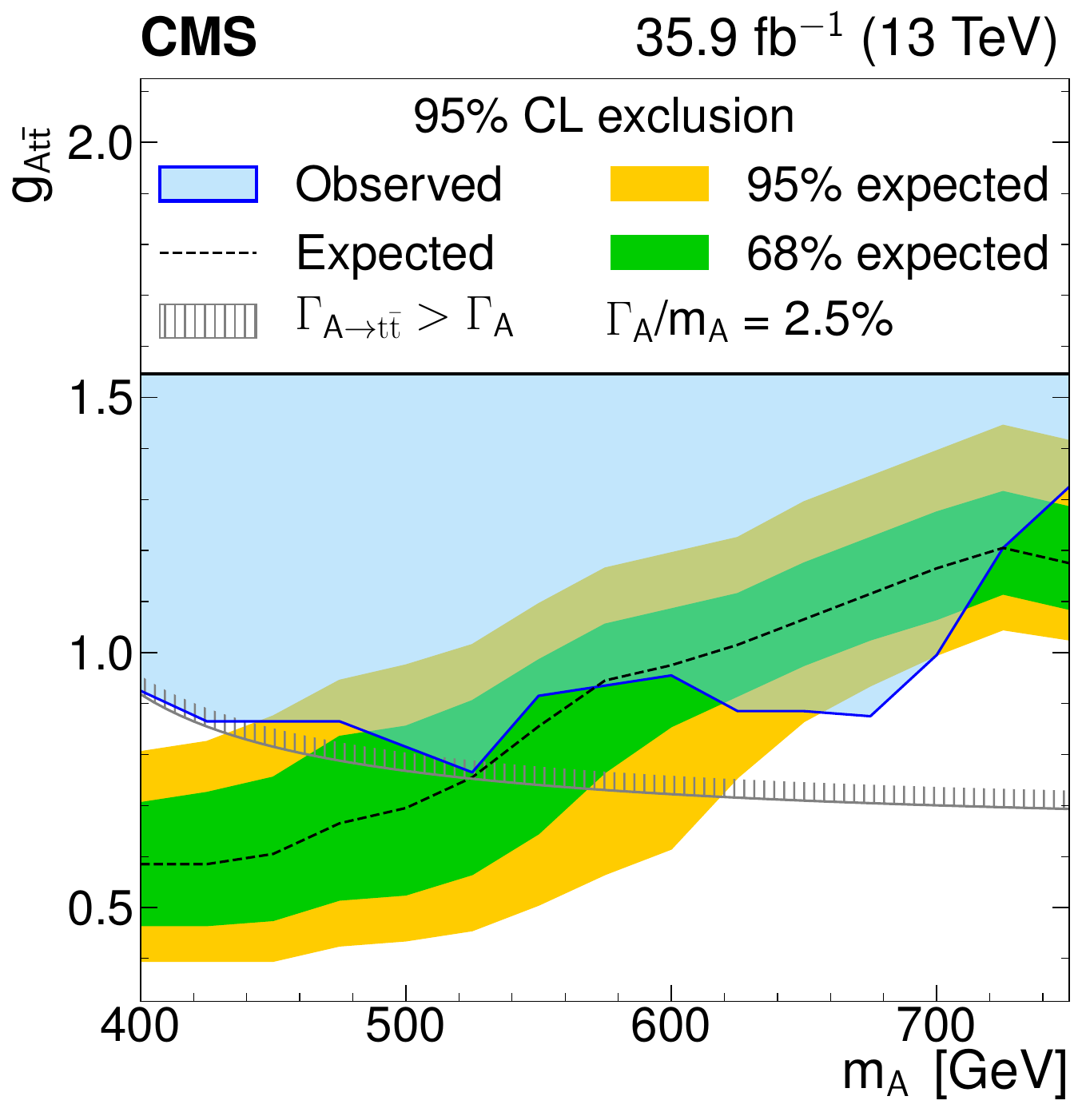}
\includegraphics[width=0.39\textwidth,keepaspectratio=true]{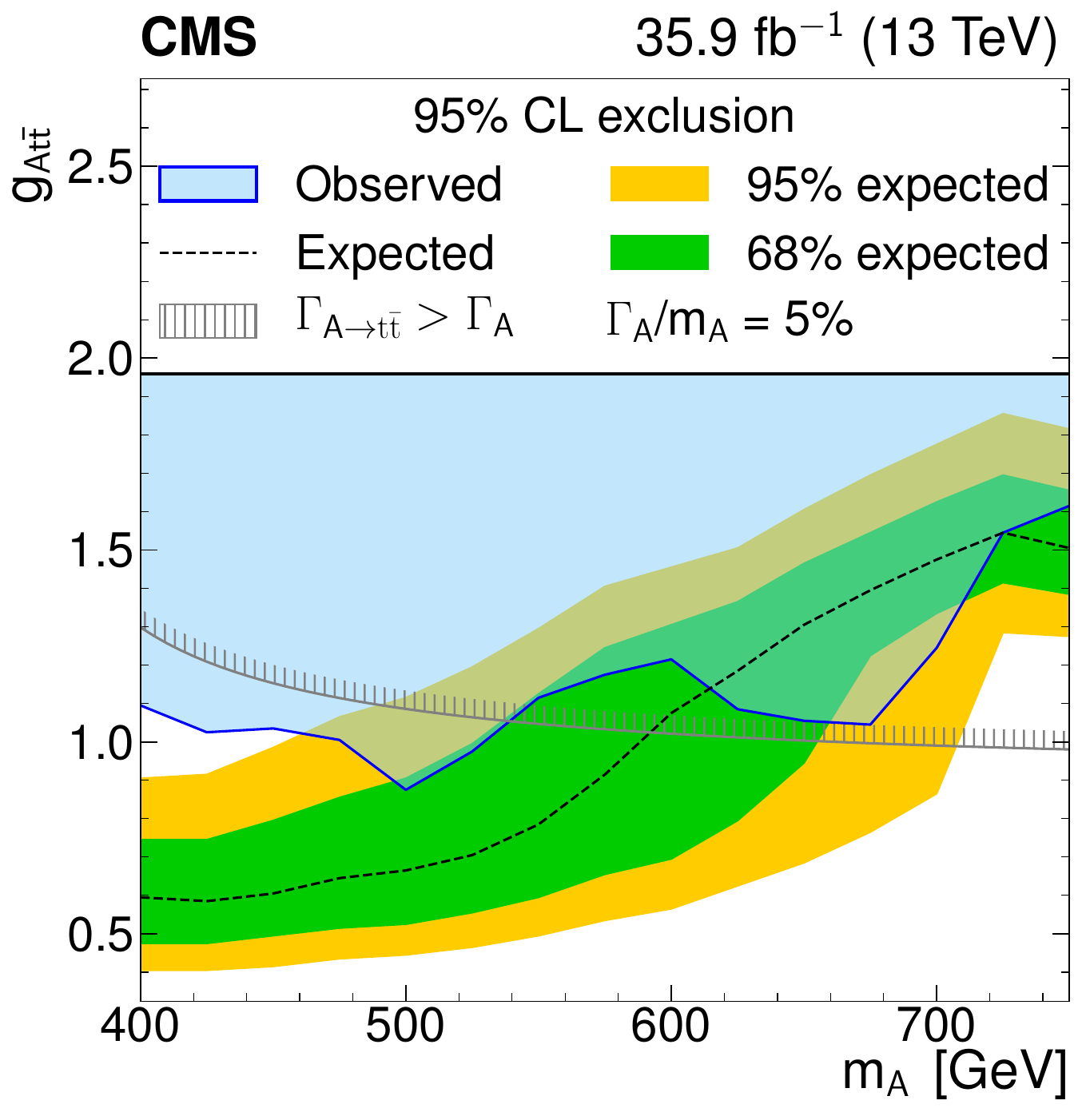}
\includegraphics[width=0.39\textwidth,keepaspectratio=true]{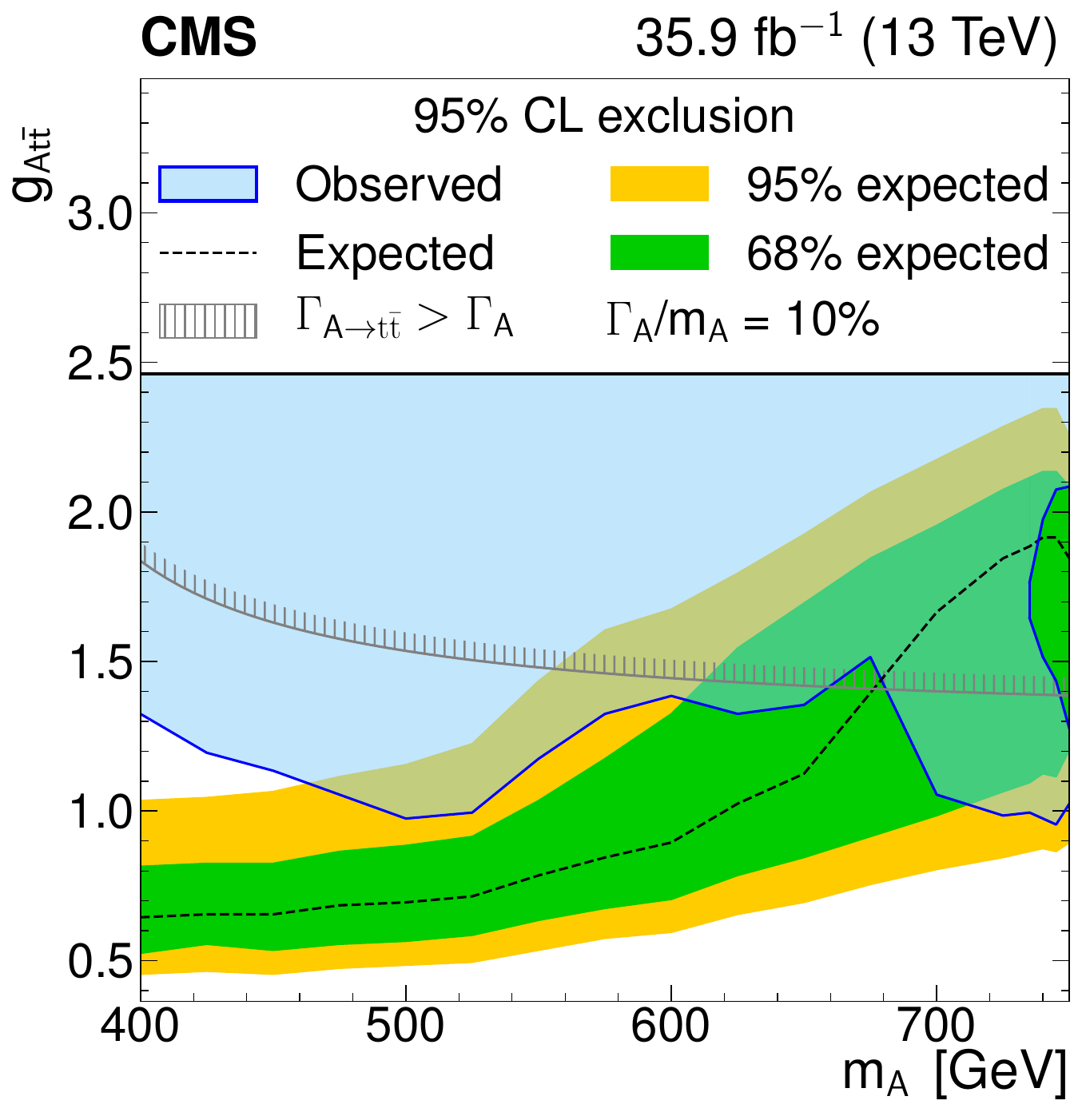}
\includegraphics[width=0.39\textwidth,keepaspectratio=true]{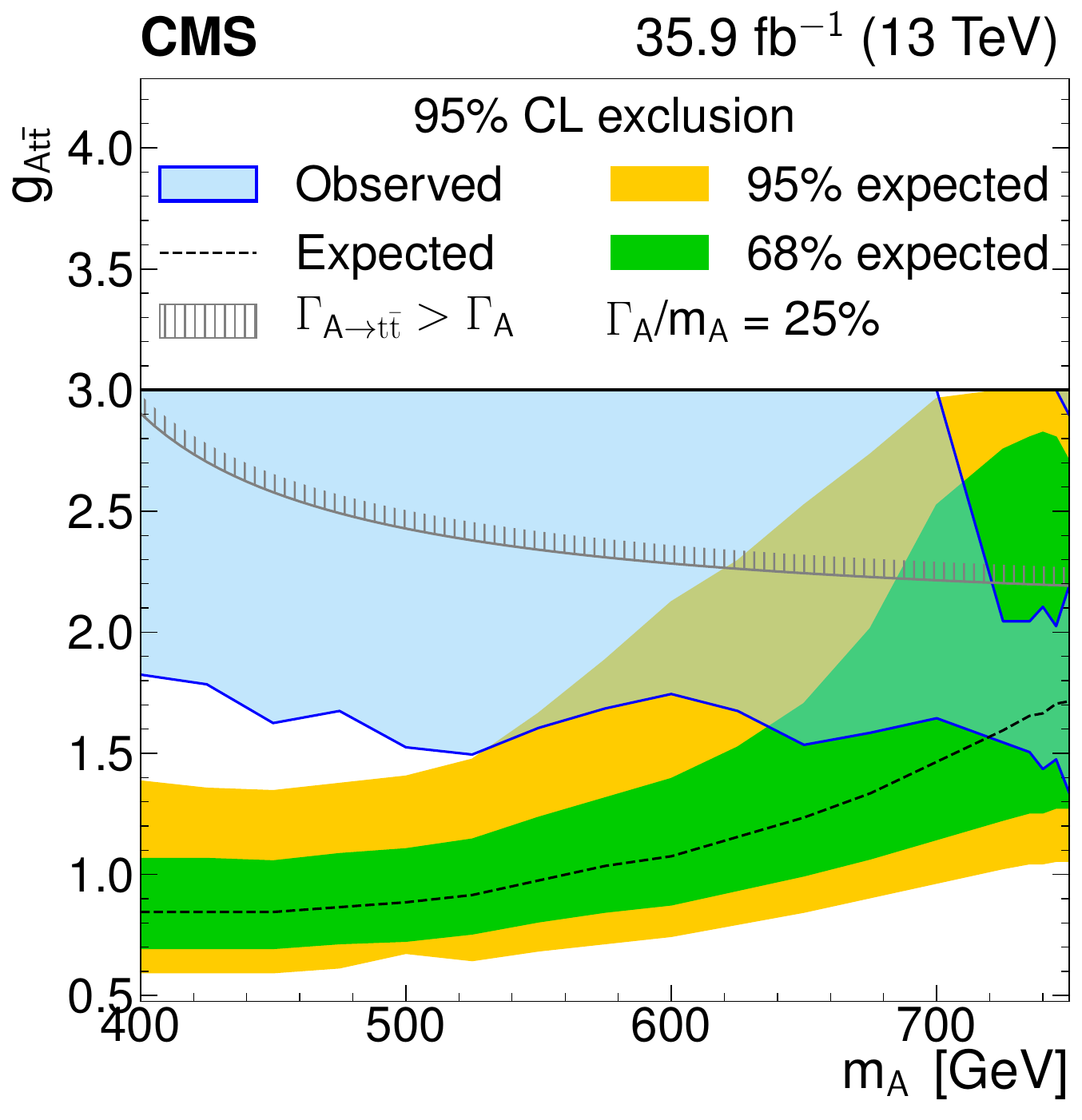}
\caption{Model-independent constraints on the coupling strength modifier as a function of the heavy pseudoscalar boson mass, for relative widths of 0.5, 1, 2.5, 5, 10, and 25\%.
The observed constraints are indicated by the blue shaded area.
The inner (green) band and the outer (yellow) band indicate the regions containing 68 and 95\%, respectively, of the distribution of constraints expected under the background-only hypothesis.
The unphysical region of phase space in which the partial width $\Gamma_{\HHOdd \to \ttbar}$ becomes larger than the total width is indicated by the hatched lines.}
\label{fig:limits_a_widths}
\end{figure}

The constraints obtained on $g_{\HH\ttbar}$ are presented in Figs.~\ref{fig:limits_h_widths} and~\ref{fig:limits_a_widths} for the scalar and the pseudoscalar scenarios, respectively.
Since the total width~$\Gamma_\HH$ is kept fixed during the scans and the partial width of $\HH \to \ttbar$ is proportional to $g_{\HH\ttbar}^2$, in some regions the partial width can exceed $\Gamma_\HH$.
These unphysical regions are marked in the figures with hatched lines.
In some cases the observed exclusion for a given mass does not extend continuously all the way to the largest probed $g_{\HH\ttbar} = 3$ (\eg, $m_{\HHEven} \approx 700$\GeV in the panel for $\Gamma_{\HHEven} / m_{\HHEven} = 25\%$ in Fig.~\ref{fig:limits_h_widths}).
This is due to the strong dependence of the shape of the signal distribution on the value of the coupling strength modifier.
For some values of $g_{\HH\ttbar}$ the shape becomes compatible with systematic variations in the background.

\begin{figure}
\centering
\includegraphics[width=0.75\textwidth,keepaspectratio=true]{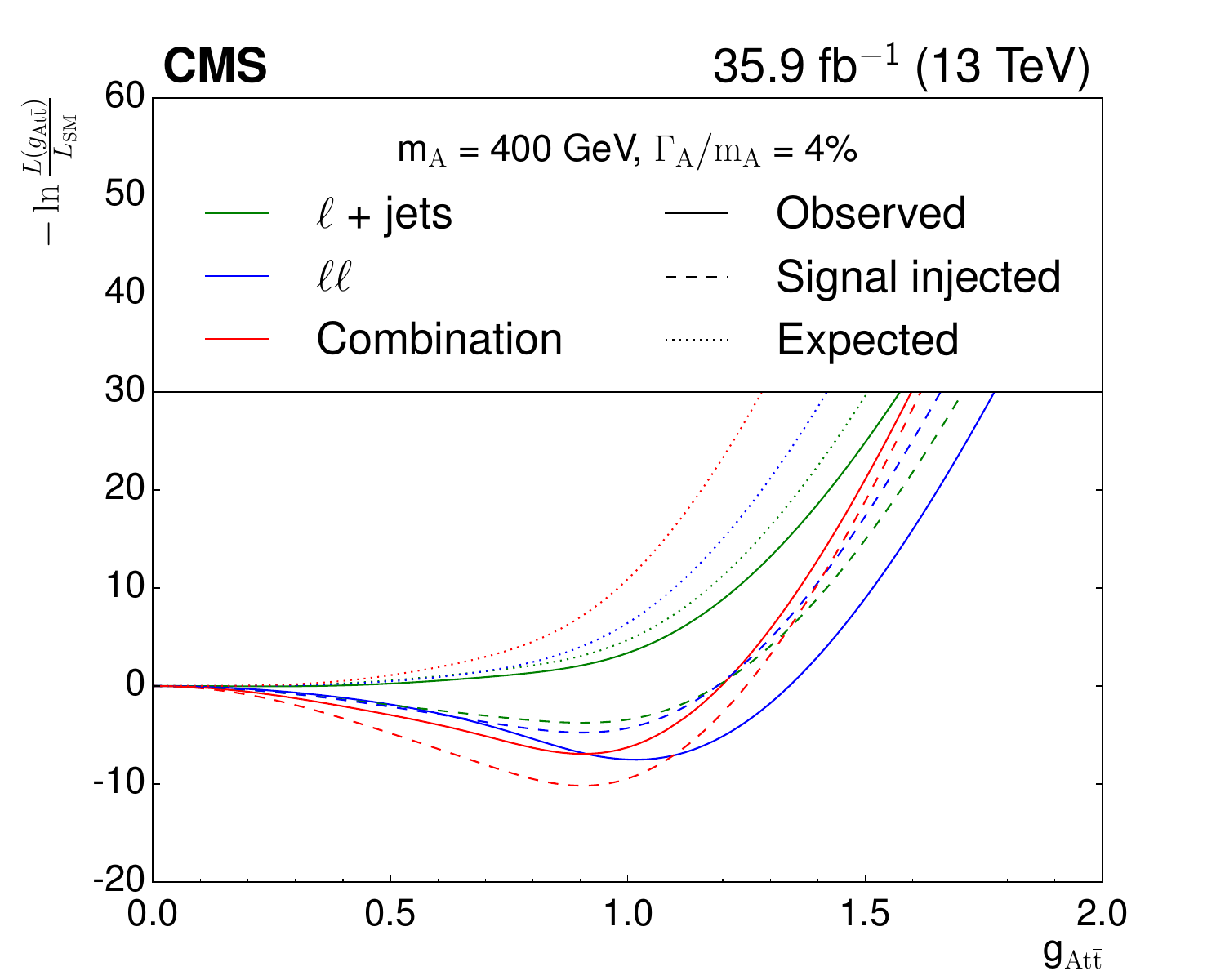}
\caption{Scans of profiled likelihood for the pseudoscalar hypothesis
with $m_\HHOdd = 400$\GeV{} and $\Gamma_\HHOdd / m_\HHOdd = 4\%$. 
The scans are shown for the single- and dilepton channels separately,
as well as for the combination. 
}
\label{fig:likelihood_scans}
\end{figure}

As evident from Fig.~\ref{fig:limits_a_widths}, there is a signal-like excess for the pseudoscalar hypotheses with low masses.
The largest deviation from the SM background is observed for a
pseudoscalar Higgs boson with a mass of 400\GeV{} and a total relative
width of 4\%, with a local significance of $3.5 \pm 0.3$ standard
deviations.
Figure~\ref{fig:likelihood_scans} shows scans of $-\ln
[L(g_{\HHOdd\ttbar}) / L_\text{SM}]$ for this hypothesis, as a
function of the coupling modifier~$g_{\HHOdd\ttbar}$. The
likelihoods~$L(g_{\HHOdd\ttbar})$ and $L_\text{SM}$ are given by
Eq.~\eqref{Eq:likelihood}. They are computed for $\mu = 1$ and 0
respectively and in both cases maximized with respect to all nuisance
parameters. The scans are shown for the observed data, as well as for
the expectations under the background-only hypothesis and in the
presence of the signal. In the latter case the coupling modifier is
set to the value obtained in the combined fit,
$g_{\HHOdd\ttbar} \approx 0.9$. In this case and in general, the
expected sensitivity is comparable between the single-lepton and
dilepton channels. The single-lepton channel is slightly more
sensitive for the scalar hypotheses and in the case of the
pseudoscalar hypotheses with larger masses. However,
Fig.~\ref{fig:likelihood_scans} demonstrates that the observed excess
is driven by the dilepton channel, which is also supported by the
comparisons between the observed and expected distributions in
Figs.~\ref{fig:SearchVars_singlelep} and~\ref{fig:SearchVars_dilep}. 

When accounting for the look-elsewhere effect~\cite{Demortier:1099967} in the mass, total width, and CP state of the heavy Higgs boson, the significance of the excess is 1.9 standard deviations, which corresponds to a $p$-value of 0.028.
We note that higher-order electroweak corrections to the SM \ttbar~production can become important in the vicinity of the pair production threshold~\cite{Kuhn:2013zoa} and may account for the excess.

\subsection{Interpretation in the hMSSM}
\label{sec:hmssmlimits}
For the hMSSM, we perform a scan over the two model parameters, $m_\HHOdd$ and $\tan\beta$.
Both \HHOdd{} and \HHEven{} bosons are included.
For each point, the coupling strength modifiers $g_{\HHOdd\ttbar}$ and $g_{\HHEven\ttbar}$, the mass $m_\HHEven$, and the widths of the two heavy Higgs bosons are determined with the 2HDMC program.
The CP-even state is more massive of the two, but the mass separation $\Delta m = m_{\HHEven} - m_{\HHOdd}$ decreases with $m_{\HHOdd}$ and $\tan\beta$.
Typical values of $\Delta m / m_{\HHOdd}$ vary from ${\approx} 20$\% for $m_{\HHOdd} = 400$\GeV, $\tan\beta = 0.5$ to ${\approx} 1$\% for $m_{\HHOdd} = 700$\GeV, $\tan\beta = 2$.
The scan is performed for $m_\HHOdd$ between 400 and 700\GeV in steps of 12.5\GeV,
and $\tan\beta$ between 0.4 and 5.0 in steps of 0.2.
Similarly to what was done above, the lower boundary $\tan\beta > 0.4$ is imposed to assure perturbative unitarity~\cite{Branco:2011iw}.
The mass and width interpolation described in Section~\ref{sec:morphing} is performed in scan points other than those corresponding to the generated signal samples.

\begin{figure}[htb!]
\centering
\includegraphics[width=0.75\textwidth,keepaspectratio=true]{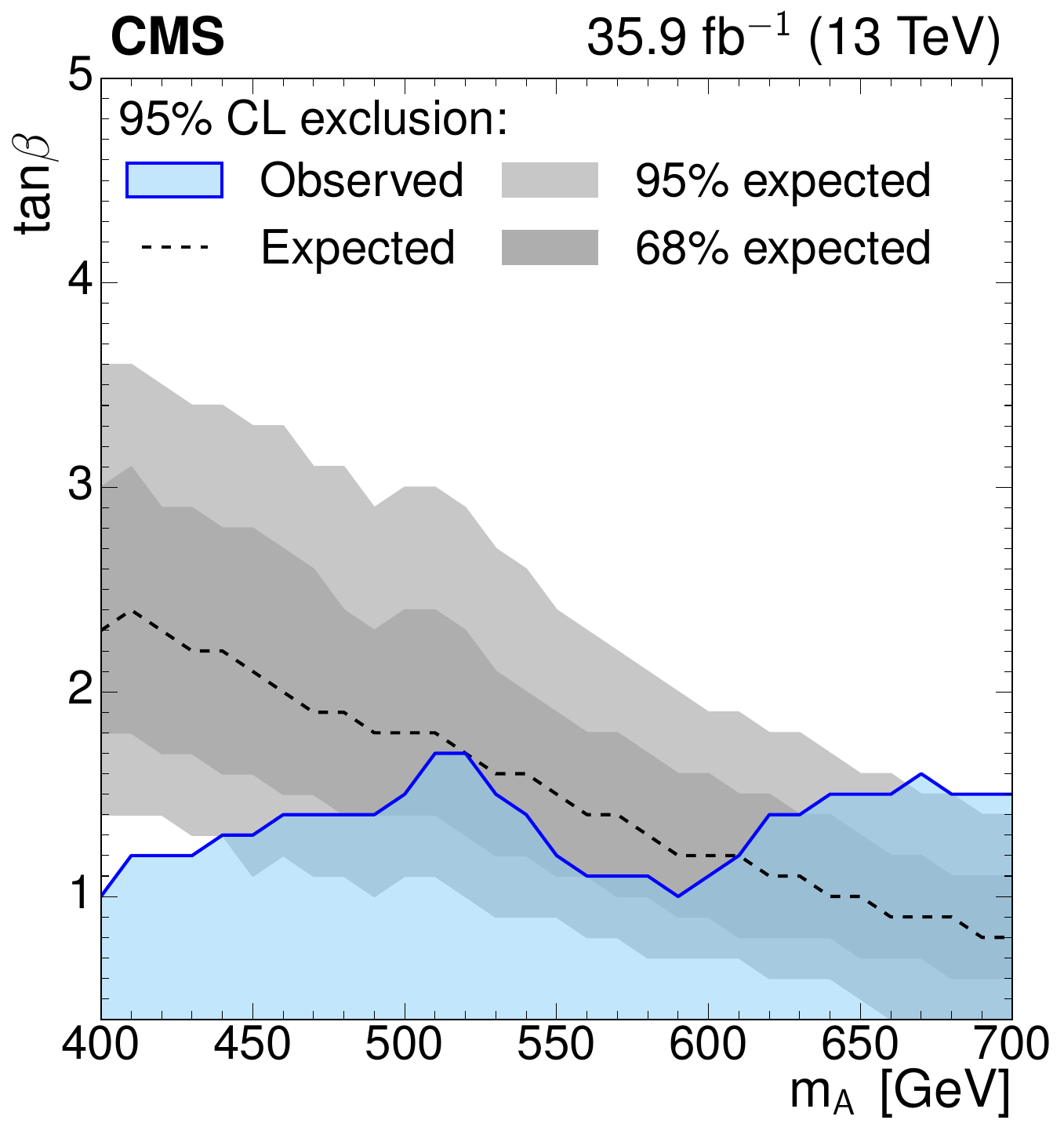}
\caption{Exclusion in the $(m_\HHOdd, \tan\beta)$ plane of the hMSSM.
  The inner (dark gray) band and the outer (light gray) band indicate the regions containing 68 and 95\%, respectively,
  of the distribution of constraints expected under the background-only hypothesis. The observed excluded region is indicated by the blue shaded area.
  Both \HHEven{} and \HHOdd{} boson signals are included with masses and widths that correspond to a given point in the plane.}
\label{fig:limits_hmssm}
\end{figure}

The expected and observed exclusions in the $(m_\HHOdd, \tan\beta)$ plane are presented in Fig.~\ref{fig:limits_hmssm}.
The upper boundary of the observed (expected) exclusion in $\tan\beta$ varies from 1.0 (2.3) at $m_\HHOdd = 400$\GeV to 1.5 (0.8) at $m_\HHOdd = 700$\GeV.
The tension between the observed exclusion and the expectation at low $m_\HHOdd$ is a manifestation of the excess discussed above.
These results can be compared to those of the search for $\PHpm \to \cPqt\cPaqb / \cPaqt\cPqb$ in Ref.~\cite{Aaboud:2018cwk}, which were also interpreted in the hMSSM benchmark, setting constraints in the $(m_{\PHpm}, \tan\beta)$ plane.
Translating the results from Ref.~\cite{Aaboud:2018cwk} in terms of $m_{\HHOdd}$, the present analysis observes a more stringent exclusion in $\tan\beta$ for $m_{\HHOdd} \approx 700$\,GeV, while the exclusion for $m_{\HHOdd} \approx 400$\,GeV is substantially weaker than in the reference due to the observed signal-like deviation.
The expected exclusion is tighter than in Ref.~\cite{Aaboud:2018cwk} throughout the considered $m_{\HHOdd}$ range.

\section{Summary}
\label{sec:summary}
Results are presented for the search for additional heavy Higgs bosons decaying to a pair of top quarks.
A data sample recorded with the CMS detector at $\sqrt{s} = 13$\TeV, corresponding to an integrated luminosity of 35.9\fbinv, is analyzed.
The final states with one or two leptons are utilized.
The invariant mass of the reconstructed \ttbar~system as well as angular variables sensitive to the spin of the new boson
are used to search for the signal, while taking into account the interference with the standard model \ttbar~production.

A moderate signal-like deviation is observed for the hypothesis of a pseudoscalar Higgs boson with the mass $m_{\HHOdd} \approx 400$\GeV.
After accounting for the look-elsewhere effect, its significance is 1.9 standard deviations.
Further improvements of the theoretical description of the standard model \ttbar~process in the vicinity of the production threshold will be needed to clarify the origin of this deviation.

Constraints on the strength of the coupling of the sought-for boson to top quarks are reported, separately for the scalar and pseudoscalar cases, for the mass ranging from 400 to 750\GeV{} and the total relative width from 0.5 to 25\%.
These are the most stringent constraints on this coupling to date.
The results are also interpreted in the hMSSM scenario in the minimal supersymmetric standard model.
This search probes the values of $m_\HHOdd$ from 400 to 700\GeV{} and excludes, at 95\% confidence level, the region with values of $\tan\beta$ below 1.0 to 1.5, depending on $m_\HHOdd$.
This extends the exclusion obtained in previous searches.

\begin{acknowledgments}
We congratulate our colleagues in the CERN accelerator departments for the excellent performance of the LHC and thank the technical and administrative staffs at CERN and at other CMS institutes for their contributions to the success of the CMS effort. In addition, we gratefully acknowledge the computing centers and personnel of the Worldwide LHC Computing Grid for delivering so effectively the computing infrastructure essential to our analyses. Finally, we acknowledge the enduring support for the construction and operation of the LHC and the CMS detector provided by the following funding agencies: BMBWF and FWF (Austria); FNRS and FWO (Belgium); CNPq, CAPES, FAPERJ, FAPERGS, and FAPESP (Brazil); MES (Bulgaria); CERN; CAS, MoST, and NSFC (China); COLCIENCIAS (Colombia); MSES and CSF (Croatia); RPF (Cyprus); SENESCYT (Ecuador); MoER, ERC IUT, PUT and ERDF (Estonia); Academy of Finland, MEC, and HIP (Finland); CEA and CNRS/IN2P3 (France); BMBF, DFG, and HGF (Germany); GSRT (Greece); NKFIA (Hungary); DAE and DST (India); IPM (Iran); SFI (Ireland); INFN (Italy); MSIP and NRF (Republic of Korea); MES (Latvia); LAS (Lithuania); MOE and UM (Malaysia); BUAP, CINVESTAV, CONACYT, LNS, SEP, and UASLP-FAI (Mexico); MOS (Montenegro); MBIE (New Zealand); PAEC (Pakistan); MSHE and NSC (Poland); FCT (Portugal); JINR (Dubna); MON, RosAtom, RAS, RFBR, and NRC KI (Russia); MESTD (Serbia); SEIDI, CPAN, PCTI, and FEDER (Spain); MOSTR (Sri Lanka); Swiss Funding Agencies (Switzerland); MST (Taipei); ThEPCenter, IPST, STAR, and NSTDA (Thailand); TUBITAK and TAEK (Turkey); NASU and SFFR (Ukraine); STFC (United Kingdom); DOE and NSF (USA).

\hyphenation{Rachada-pisek} Individuals have received support from the Marie-Curie program and the European Research Council and Horizon 2020 Grant, contract Nos.\ 675440, 752730, and 765710 (European Union); the Leventis Foundation; the A.P.\ Sloan Foundation; the Alexander von Humboldt Foundation; the Belgian Federal Science Policy Office; the Fonds pour la Formation \`a la Recherche dans l'Industrie et dans l'Agriculture (FRIA-Belgium); the Agentschap voor Innovatie door Wetenschap en Technologie (IWT-Belgium); the F.R.S.-FNRS and FWO (Belgium) under the ``Excellence of Science -- EOS" -- be.h project n.\ 30820817; the Beijing Municipal Science \& Technology Commission, No. Z181100004218003; the Ministry of Education, Youth and Sports (MEYS) of the Czech Republic; the Lend\"ulet (``Momentum") Program and the J\'anos Bolyai Research Scholarship of the Hungarian Academy of Sciences, the New National Excellence Program \'UNKP, the NKFIA research grants 123842, 123959, 124845, 124850, 125105, 128713, 128786, and 129058 (Hungary); the Council of Science and Industrial Research, India; the HOMING PLUS program of the Foundation for Polish Science, cofinanced from European Union, Regional Development Fund, the Mobility Plus program of the Ministry of Science and Higher Education, the National Science Center (Poland), contracts Harmonia 2014/14/M/ST2/00428, Opus 2014/13/B/ST2/02543, 2014/15/B/ST2/03998, and 2015/19/B/ST2/02861, Sonata-bis 2012/07/E/ST2/01406; the National Priorities Research Program by Qatar National Research Fund; the Ministry of Science and Education, grant no. 3.2989.2017 (Russia); the Programa Estatal de Fomento de la Investigaci{\'o}n Cient{\'i}fica y T{\'e}cnica de Excelencia Mar\'{\i}a de Maeztu, grant MDM-2015-0509 and the Programa Severo Ochoa del Principado de Asturias; the Thalis and Aristeia programs cofinanced by EU-ESF and the Greek NSRF; the Rachadapisek Sompot Fund for Postdoctoral Fellowship, Chulalongkorn University and the Chulalongkorn Academic into Its 2nd Century Project Advancement Project (Thailand); the Welch Foundation, contract C-1845; and the Weston Havens Foundation (USA).
\end{acknowledgments}

\bibliography{auto_generated}

\providecommand{\href}[2]{#2}\begingroup\raggedright\begin{thebibliography}{10}%
\makeatletter
\providecommand{\hrefCMSnoop }[0]{\@secondoftwo}%
\makeatother
\providecommand{\doi}{\texttt{doi:}\begingroup \urlstyle{tt}\Url}

\bibitem{Aad:2012tfa}
\hrefCMSnoop {}{{ATLAS Collaboration}, ``Observation of a new particle in the
  search for the standard model {Higgs} boson with the {ATLAS} detector at the
  {LHC}'',} \textit{ Phys. Lett. B} \textbf{ 716} (2012) 1,
  \href{http://dx.doi.org/10.1016/j.physletb.2012.08.020}{\doi{10.1016/j.physletb.2012.08.020}},
\href{http://www.arXiv.org/abs/1207.7214}{\texttt{arXiv:1207.7214}}.

\bibitem{Chatrchyan:2012xdj}
\hrefCMSnoop {}{{CMS Collaboration}, ``Observation of a new boson at a mass of
  125\,{GeV} with the {CMS} experiment at the {LHC}'',} \textit{ Phys. Lett. B}
  \textbf{ 716} (2012) 30,
  \href{http://dx.doi.org/10.1016/j.physletb.2012.08.021}{\doi{10.1016/j.physletb.2012.08.021}},
\href{http://www.arXiv.org/abs/1207.7235}{\texttt{arXiv:1207.7235}}.

\bibitem{Chatrchyan:2013lba}
\hrefCMSnoop {}{{CMS Collaboration}, ``Observation of a new boson with mass
  near 125\,{GeV} in pp~collisions at $\sqrt{s} = 7$ and 8\,{TeV}'',} \textit{
  JHEP} \textbf{ 06} (2013) 081,
  \href{http://dx.doi.org/10.1007/JHEP06(2013)081}{\doi{10.1007/JHEP06(2013)081}},
\href{http://www.arXiv.org/abs/1303.4571}{\texttt{arXiv:1303.4571}}.

\bibitem{Branco:2011iw}
G.~C. Branco\hrefCMSnoop {}{ { et~al.}, ``Theory and phenomenology of
  two-{H}iggs-doublet models'',} \textit{ Phys. Rept.} \textbf{ 516} (2012) 1,
  \href{http://dx.doi.org/10.1016/j.physrep.2012.02.002}{\doi{10.1016/j.physrep.2012.02.002}},
\href{http://www.arXiv.org/abs/1106.0034}{\texttt{arXiv:1106.0034}}.

\bibitem{Wess:1974tw}
\hrefCMSnoop {}{J.~Wess and B.~Zumino, ``Supergauge transformations in
  four-dimensions'',} \textit{ Nucl. Phys. B} \textbf{ 70} (1974) 39,
\href{http://dx.doi.org/10.1016/0550-3213(74)90355-1}{\doi{10.1016/0550-3213(74)90355-1}}.

\bibitem{Dimopoulos:1981zb}
\hrefCMSnoop {}{S.~Dimopoulos and H.~Georgi, ``Softly broken supersymmetry and
  {SU(5)}'',} \textit{ Nucl. Phys. B} \textbf{ 193} (1981) 150,
\href{http://dx.doi.org/10.1016/0550-3213(81)90522-8}{\doi{10.1016/0550-3213(81)90522-8}}.

\bibitem{Huitu:2019}
K.~Huitu\hrefCMSnoop {}{ { et~al.}, ``Probing pseudo-{G}oldstone dark matter at
  the {LHC}'',} \textit{ Phys. Rev. D} \textbf{ 100} (2019) 015009,
  \href{http://dx.doi.org/10.1103/PhysRevD.100.015009}{\doi{10.1103/PhysRevD.100.015009}},
\href{http://www.arXiv.org/abs/1812.05952}{\texttt{arXiv:1812.05952}}.

\bibitem{Muhlleitner:2017dkd}
\hrefCMSnoop {}{M.~M{\"u}hlleitner, M.~O.~P. Sampaio, R.~Santos, and
  J.~Wittbrodt, ``Phenomenological comparison of models with extended {H}iggs
  sectors'',} \textit{ JHEP} \textbf{ 08} (2017) 132,
  \href{http://dx.doi.org/10.1007/JHEP08(2017)132}{\doi{10.1007/JHEP08(2017)132}},
\href{http://www.arXiv.org/abs/1703.07750}{\texttt{arXiv:1703.07750}}.

\bibitem{DMLHC:2015}
\hrefCMSnoop {}{J.~Abdallah { et~al.}, ``Simplified models for dark matter
  searches at the {LHC}'',} \textit{ Phys. Dark Univ.} \textbf{ 9--10} (2015)
  8,
  \href{http://dx.doi.org/10.1016/j.dark.2015.08.001}{\doi{10.1016/j.dark.2015.08.001}},
\href{http://www.arXiv.org/abs/1506.03116}{\texttt{arXiv:1506.03116}}.

\bibitem{Arina:2016}
C.~Arina\hrefCMSnoop {}{ { et~al.}, ``A comprehensive approach to dark matter
  studies: exploration of simplified top-philic models'',} \textit{ JHEP}
  \textbf{ 11} (2016) 111,
  \href{http://dx.doi.org/10.1007/JHEP11(2016)111}{\doi{10.1007/JHEP11(2016)111}},
\href{http://www.arXiv.org/abs/1605.09242}{\texttt{arXiv:1605.09242}}.

\bibitem{Craig:2013hca}
\hrefCMSnoop {}{N.~Craig, J.~Galloway, and S.~Thomas, ``Searching for signs of
  the second {H}iggs doublet'',} 2013.
\href{http://www.arXiv.org/abs/1305.2424}{\texttt{arXiv:1305.2424}}.

\bibitem{Lane:2019}
\hrefCMSnoop {}{K.~Lane and W.~Shepherd, ``Natural stabilization of the {H}iggs
  boson's mass and alignment'',} \textit{ Phys. Rev. D} \textbf{ 99} (2019)
  055015,
  \href{http://dx.doi.org/10.1103/PhysRevD.99.055015}{\doi{10.1103/PhysRevD.99.055015}},
\href{http://www.arXiv.org/abs/1808.07927}{\texttt{arXiv:1808.07927}}.

\bibitem{Djouadi:2013uqa}
A.~Djouadi\hrefCMSnoop {}{ { et~al.}, ``The post-{Higgs} {MSSM} scenario:
  {Habemus MSSM}?'',} \textit{ Eur. Phys. J. C} \textbf{ 73} (2013) 2650,
  \href{http://dx.doi.org/10.1140/epjc/s10052-013-2650-0}{\doi{10.1140/epjc/s10052-013-2650-0}},
\href{http://www.arXiv.org/abs/1307.5205}{\texttt{arXiv:1307.5205}}.

\bibitem{Gaemers:1984sj}
\hrefCMSnoop {}{K.~J.~F. Gaemers and F.~Hoogeveen, ``Higgs production and decay
  into heavy flavours with the gluon fusion mechanism'',} \textit{ Phys. Lett.
  B} \textbf{ 146} (1984) 347,
\href{http://dx.doi.org/10.1016/0370-2693(84)91711-8}{\doi{10.1016/0370-2693(84)91711-8}}.

\bibitem{Dicus:1994bm}
\hrefCMSnoop {}{D.~Dicus, A.~Stange, and S.~Willenbrock, ``Higgs decay to top
  quarks at hadron colliders'',} \textit{ Phys. Lett. B} \textbf{ 333} (1994)
  126,
  \href{http://dx.doi.org/10.1016/0370-2693(94)91017-0}{\doi{10.1016/0370-2693(94)91017-0}},
\href{http://www.arXiv.org/abs/hep-ph/9404359}{\texttt{arXiv:hep-ph/9404359}}.

\bibitem{Bernreuther:1997gs}
\hrefCMSnoop {}{W.~Bernreuther, M.~Flesch, and P.~Haberl, ``Signatures of
  {H}iggs bosons in the top quark decay channel at hadron colliders'',}
  \textit{ Phys. Rev. D} \textbf{ 58} (1998) 114031,
  \href{http://dx.doi.org/10.1103/PhysRevD.58.114031}{\doi{10.1103/PhysRevD.58.114031}},
\href{http://www.arXiv.org/abs/hep-ph/9709284}{\texttt{arXiv:hep-ph/9709284}}.

\bibitem{Carena:2016}
\hrefCMSnoop {}{M.~Carena and Z.~Liu, ``Challenges and opportunities for heavy
  scalar searches in the \ttbar~channel at the {LHC}'',} \textit{ JHEP}
  \textbf{ 11} (2016) 159,
  \href{http://dx.doi.org/10.1007/JHEP11(2016)159}{\doi{10.1007/JHEP11(2016)159}},
\href{http://www.arXiv.org/abs/1608.07282}{\texttt{arXiv:1608.07282}}.

\bibitem{Djouadi:2019}
\hrefCMSnoop {}{A.~Djouadi, J.~Ellis, A.~Popov, and J.~Quevillon,
  ``Interference effects in \ttbar~production at the {LHC} as a window on new
  physics'',} \textit{ JHEP} \textbf{ 03} (2019) 119,
  \href{http://dx.doi.org/10.1007/JHEP03(2019)119}{\doi{10.1007/JHEP03(2019)119}},
\href{http://www.arXiv.org/abs/1901.03417}{\texttt{arXiv:1901.03417}}.

\bibitem{atlashttottbar8TeV}
\hrefCMSnoop {}{{ATLAS Collaboration}, ``Search for heavy {H}iggs bosons {A/H}
  decaying to a top quark pair in pp~collisions at $\sqrt{s} = 8$\,{TeV} with
  the {ATLAS} detector'',} \textit{ Phys. Rev. Lett.} \textbf{ 119} (2017)
  191803,
  \href{http://dx.doi.org/10.1103/PhysRevLett.119.191803}{\doi{10.1103/PhysRevLett.119.191803}},
\href{http://www.arXiv.org/abs/1707.06025}{\texttt{arXiv:1707.06025}}.

\bibitem{cmsSUSYsamesign13TeV}
\hrefCMSnoop {}{{CMS Collaboration}, ``Search for physics beyond the standard
  model in events with two leptons of same sign, missing transverse momentum,
  and jets in proton-proton collisions at $\sqrt{s}$ = 13\,{TeV}'',} \textit{
  Eur. Phys. J. C} \textbf{ 77} (2017) 578,
  \href{http://dx.doi.org/10.1140/epjc/s10052-017-5079-z}{\doi{10.1140/epjc/s10052-017-5079-z}},
\href{http://www.arXiv.org/abs/1704.07323}{\texttt{arXiv:1704.07323}}.

\bibitem{atlasZprime13TeV}
\hrefCMSnoop {}{{ATLAS Collaboration}, ``Search for heavy particles decaying
  into top-quark pairs using lepton-plus-jets events in proton--proton
  collisions at $\sqrt{s} = 13$\,{TeV} with the {ATLAS} detector'',} \textit{
  Eur. Phys. J. C} \textbf{ 78} (2018) 565,
  \href{http://dx.doi.org/10.1140/epjc/s10052-018-5995-6}{\doi{10.1140/epjc/s10052-018-5995-6}},
\href{http://www.arXiv.org/abs/1804.10823}{\texttt{arXiv:1804.10823}}.

\bibitem{cmsZprime13TeV}
\hrefCMSnoop {}{{CMS Collaboration}, ``Search for $\ttbar$ resonances in highly
  boosted $\text{lepton}+\text{jets}$ and fully hadronic final states in
  proton-proton collisions at $\sqrt{s}$ = 13\,{TeV}'',} \textit{ JHEP}
  \textbf{ 07} (2017) 001,
  \href{http://dx.doi.org/10.1007/JHEP07(2017)001}{\doi{10.1007/JHEP07(2017)001}},
\href{http://www.arXiv.org/abs/1704.03366}{\texttt{arXiv:1704.03366}}.

\bibitem{Aaboud:2018cwk}
\hrefCMSnoop {}{{ATLAS Collaboration}, ``Search for charged {H}iggs bosons
  decaying into top and bottom quarks at $\sqrt{s} = 13$\,{TeV} with the
  {ATLAS} detector'',} \textit{ JHEP} \textbf{ 11} (2018) 085,
  \href{http://dx.doi.org/10.1007/JHEP11(2018)085}{\doi{10.1007/JHEP11(2018)085}},
\href{http://www.arXiv.org/abs/1808.03599}{\texttt{arXiv:1808.03599}}.

\bibitem{Khachatryan:2015qxa}
\hrefCMSnoop {}{{CMS Collaboration}, ``Search for a charged {H}iggs boson in pp
  collisions at $\sqrt{s} = 8$\,{TeV}'',} \textit{ JHEP} \textbf{ 11} (2015)
  018,
  \href{http://dx.doi.org/10.1007/JHEP11(2015)018}{\doi{10.1007/JHEP11(2015)018}},
\href{http://www.arXiv.org/abs/1508.07774}{\texttt{arXiv:1508.07774}}.

\bibitem{Chatrchyan:2008zzk}
\hrefCMSnoop {}{{CMS Collaboration}, ``The {CMS} experiment at the {CERN}
  {LHC}'',} \textit{ JINST} \textbf{ 3} (2008) S08004,
\href{http://dx.doi.org/10.1088/1748-0221/3/08/S08004}{\doi{10.1088/1748-0221/3/08/S08004}}.

\bibitem{Cacciari:2008gp}
\hrefCMSnoop {}{M.~Cacciari, G.~P. Salam, and G.~Soyez, ``The anti-$k_\text{T}$
  jet clustering algorithm'',} \textit{ JHEP} \textbf{ 04} (2008) 063,
  \href{http://dx.doi.org/10.1088/1126-6708/2008/04/063}{\doi{10.1088/1126-6708/2008/04/063}},
\href{http://www.arXiv.org/abs/0802.1189}{\texttt{arXiv:0802.1189}}.

\bibitem{Cacciari:2011ma}
\hrefCMSnoop {}{M.~Cacciari, G.~P. Salam, and G.~Soyez, ``{FastJet} user
  manual'',} \textit{ Eur. Phys. J. C} \textbf{ 72} (2012) 1896,
  \href{http://dx.doi.org/10.1140/epjc/s10052-012-1896-2}{\doi{10.1140/epjc/s10052-012-1896-2}},
\href{http://www.arXiv.org/abs/1111.6097}{\texttt{arXiv:1111.6097}}.

\bibitem{CMS-PRF-14-001}
\hrefCMSnoop {}{{CMS Collaboration}, ``Particle-flow reconstruction and global
  event description with the {CMS} detector'',} \textit{ JINST} \textbf{ 12}
  (2017) P10003,
  \href{http://dx.doi.org/10.1088/1748-0221/12/10/P10003}{\doi{10.1088/1748-0221/12/10/P10003}},
\href{http://www.arXiv.org/abs/1706.04965}{\texttt{arXiv:1706.04965}}.

\bibitem{Khachatryan:2016kdb}
\hrefCMSnoop {}{{CMS Collaboration}, ``Jet energy scale and resolution in the
  {CMS} experiment in pp~collisions at 8\,{TeV}'',} \textit{ JINST} \textbf{
  12} (2017) P02014,
  \href{http://dx.doi.org/10.1088/1748-0221/12/02/P02014}{\doi{10.1088/1748-0221/12/02/P02014}},
\href{http://www.arXiv.org/abs/1607.03663}{\texttt{arXiv:1607.03663}}.

\bibitem{PAS-JME-16-003}
\href {https://cds.cern.ch/record/2256875}{{CMS Collaboration}, ``Jet
  algorithms performance in 13\,{TeV} data'',} CMS Physics Analysis Summary
  CMS-PAS-JME-16-003, 2017.

\bibitem{CMS-BTV-16-002}
\hrefCMSnoop {}{{CMS Collaboration}, ``Identification of heavy-flavour jets
  with the {CMS} detector in pp~collisions at 13\,{TeV}'',} \textit{ JINST}
  \textbf{ 13} (2018) P05011,
  \href{http://dx.doi.org/10.1088/1748-0221/13/05/P05011}{\doi{10.1088/1748-0221/13/05/P05011}},
\href{http://www.arXiv.org/abs/1712.07158}{\texttt{arXiv:1712.07158}}.

\bibitem{Khachatryan:2015hwa}
\hrefCMSnoop {}{{CMS Collaboration}, ``Performance of electron reconstruction
  and selection with the {CMS} detector in proton-proton collisions at
  $\sqrt{s} = 8$\,{\TeV}'',} \textit{ JINST} \textbf{ 10} (2015) P06005,
  \href{http://dx.doi.org/10.1088/1748-0221/10/06/P06005}{\doi{10.1088/1748-0221/10/06/P06005}},
\href{http://www.arXiv.org/abs/1502.02701}{\texttt{arXiv:1502.02701}}.

\bibitem{Sirunyan:2018fpa}
\hrefCMSnoop {}{{CMS Collaboration}, ``Performance of the {CMS} muon detector
  and muon reconstruction with proton-proton collisions at $\sqrt{s}=
  13$\,{TeV}'',} \textit{ JINST} \textbf{ 13} (2018) P06015,
  \href{http://dx.doi.org/10.1088/1748-0221/13/06/P06015}{\doi{10.1088/1748-0221/13/06/P06015}},
\href{http://www.arXiv.org/abs/1804.04528}{\texttt{arXiv:1804.04528}}.

\bibitem{Sirunyan:2019kia}
\hrefCMSnoop {}{{CMS Collaboration}, ``Performance of missing transverse
  momentum reconstruction in proton-proton collisions at $\sqrt{s} = 13$\,{TeV}
  using the {CMS} detector'',} \textit{ JINST} \textbf{ 14} (2019) P07004,
  \href{http://dx.doi.org/10.1088/1748-0221/14/07/P07004}{\doi{10.1088/1748-0221/14/07/P07004}},
\href{http://www.arXiv.org/abs/1903.06078}{\texttt{arXiv:1903.06078}}.

\bibitem{CMS-PAS-LUM-17-001}
\href {https://cds.cern.ch/record/2257069}{{CMS Collaboration}, ``{CMS}
  luminosity measurements for the 2016 data taking period'',} CMS Physics
  Analysis Summary CMS-PAS-LUM-17-001, 2017.

\bibitem{Khachatryan:2016bia}
\hrefCMSnoop {}{{CMS Collaboration}, ``The {CMS} trigger system'',} \textit{
  JINST} \textbf{ 12} (2017) P01020,
  \href{http://dx.doi.org/10.1088/1748-0221/12/01/P01020}{\doi{10.1088/1748-0221/12/01/P01020}},
\href{http://www.arXiv.org/abs/1609.02366}{\texttt{arXiv:1609.02366}}.

\bibitem{Alwall:2014hca}
J.~Alwall\hrefCMSnoop {}{ { et~al.}, ``{The automated computation of tree-level
  and next-to-leading order differential cross sections, and their matching to
  parton shower simulations}'',} \textit{ JHEP} \textbf{ 07} (2014) 079,
  \href{http://dx.doi.org/10.1007/JHEP07(2014)079}{\doi{10.1007/JHEP07(2014)079}},
\href{http://www.arXiv.org/abs/1405.0301}{\texttt{arXiv:1405.0301}}.

\bibitem{Spira:1995rr}
\hrefCMSnoop {}{M.~Spira, A.~Djouadi, D.~Graudenz, and P.~M. Zerwas, ``{H}iggs
  boson production at the {LHC}'',} \textit{ Nucl. Phys. B} \textbf{ 453}
  (1995) 17,
  \href{http://dx.doi.org/10.1016/0550-3213(95)00379-7}{\doi{10.1016/0550-3213(95)00379-7}},
\href{http://www.arXiv.org/abs/hep-ph/9504378}{\texttt{arXiv:hep-ph/9504378}}.

\bibitem{Ball:2014uwa}
\hrefCMSnoop {}{{NNPDF} Collaboration, ``Parton distributions for the {LHC}
  {Run} {II}'',} \textit{ JHEP} \textbf{ 04} (2015) 040,
  \href{http://dx.doi.org/10.1007/JHEP04(2015)040}{\doi{10.1007/JHEP04(2015)040}},
\href{http://www.arXiv.org/abs/1410.8849}{\texttt{arXiv:1410.8849}}.

\bibitem{Sjostrand:2014zea}
T.~Sj{\"o}strand\hrefCMSnoop {}{ { et~al.}, ``An introduction to {PYTHIA
  8.2}'',} \textit{ Comput. Phys. Commun.} \textbf{ 191} (2015) 159,
  \href{http://dx.doi.org/10.1016/j.cpc.2015.01.024}{\doi{10.1016/j.cpc.2015.01.024}},
\href{http://www.arXiv.org/abs/1410.3012}{\texttt{arXiv:1410.3012}}.

\bibitem{Skands:2014pea}
\hrefCMSnoop {}{P.~Skands, S.~Carrazza, and J.~Rojo, ``Tuning {PYTHIA} 8.1: the
  {Monash} 2013 tune'',} \textit{ Eur. Phys. J. C} \textbf{ 74} (2014) 3024,
  \href{http://dx.doi.org/10.1140/epjc/s10052-014-3024-y}{\doi{10.1140/epjc/s10052-014-3024-y}},
\href{http://www.arXiv.org/abs/1404.5630}{\texttt{arXiv:1404.5630}}.

\bibitem{Khachatryan:2015pea}
\hrefCMSnoop {}{{CMS Collaboration}, ``Event generator tunes obtained from
  underlying event and multiparton scattering measurements'',} \textit{ Eur.
  Phys. J. C} \textbf{ 76} (2016) 155,
  \href{http://dx.doi.org/10.1140/epjc/s10052-016-3988-x}{\doi{10.1140/epjc/s10052-016-3988-x}},
\href{http://www.arXiv.org/abs/1512.00815}{\texttt{arXiv:1512.00815}}.

\bibitem{Hespel:2016qaf}
\hrefCMSnoop {}{B.~Hespel, F.~Maltoni, and E.~Vryonidou, ``Signal background
  interference effects in heavy scalar production and decay to a top-anti-top
  pair'',} \textit{ JHEP} \textbf{ 10} (2016) 016,
  \href{http://dx.doi.org/10.1007/JHEP10(2016)016}{\doi{10.1007/JHEP10(2016)016}},
\href{http://www.arXiv.org/abs/1606.04149}{\texttt{arXiv:1606.04149}}.

\bibitem{Harlander:2012pb}
\hrefCMSnoop {}{R.~V. Harlander, S.~Liebler, and H.~Mantler, ``{SusHi}: A
  program for the calculation of {Higgs} production in gluon fusion and
  bottom-quark annihilation in the standard model and the {MSSM}'',} \textit{
  Comput. Phys. Commun.} \textbf{ 184} (2013) 1605,
  \href{http://dx.doi.org/10.1016/j.cpc.2013.02.006}{\doi{10.1016/j.cpc.2013.02.006}},
\href{http://www.arXiv.org/abs/1212.3249}{\texttt{arXiv:1212.3249}}.

\bibitem{Eriksson:2009ws}
\hrefCMSnoop {}{D.~Eriksson, J.~Rathsman, and O.~Stal, ``{2HDMC}:
  two-{H}iggs-doublet model calculator physics and manual'',} \textit{ Comput.
  Phys. Commun.} \textbf{ 181} (2010) 189,
  \href{http://dx.doi.org/10.1016/j.cpc.2009.09.011}{\doi{10.1016/j.cpc.2009.09.011}},
\href{http://www.arXiv.org/abs/0902.0851}{\texttt{arXiv:0902.0851}}.

\bibitem{Nason:2004rx}
\hrefCMSnoop {}{P.~Nason, ``A new method for combining {NLO} {QCD} with shower
  {Monte Carlo} algorithms'',} \textit{ JHEP} \textbf{ 11} (2004) 040,
  \href{http://dx.doi.org/10.1088/1126-6708/2004/11/040}{\doi{10.1088/1126-6708/2004/11/040}},
\href{http://www.arXiv.org/abs/hep-ph/0409146}{\texttt{arXiv:hep-ph/0409146}}.

\bibitem{Frixione:2007vw}
\hrefCMSnoop {}{S.~Frixione, P.~Nason, and C.~Oleari, ``Matching {NLO} {QCD}
  computations with parton shower simulations: the {POWHEG} method'',} \textit{
  JHEP} \textbf{ 11} (2007) 070,
  \href{http://dx.doi.org/10.1088/1126-6708/2007/11/070}{\doi{10.1088/1126-6708/2007/11/070}},
\href{http://www.arXiv.org/abs/0709.2092}{\texttt{arXiv:0709.2092}}.

\bibitem{Alioli:2010xd}
\hrefCMSnoop {}{S.~Alioli, P.~Nason, C.~Oleari, and E.~Re, ``A general
  framework for implementing {NLO} calculations in shower {M}onte {C}arlo
  programs: the {POWHEG} {BOX}'',} \textit{ JHEP} \textbf{ 06} (2010) 043,
  \href{http://dx.doi.org/10.1007/JHEP06(2010)043}{\doi{10.1007/JHEP06(2010)043}},
\href{http://www.arXiv.org/abs/1002.2581}{\texttt{arXiv:1002.2581}}.

\bibitem{Campbell:2014kua}
\hrefCMSnoop {}{J.~M. Campbell, R.~K. Ellis, P.~Nason, and E.~Re, ``Top-pair
  production and decay at {NLO} matched with parton showers'',} \textit{ JHEP}
  \textbf{ 04} (2015) 114,
  \href{http://dx.doi.org/10.1007/JHEP04(2015)114}{\doi{10.1007/JHEP04(2015)114}},
\href{http://www.arXiv.org/abs/1412.1828}{\texttt{arXiv:1412.1828}}.

\bibitem{CMS-PAS-TOP-16-021}
\href {https://cds.cern.ch/record/2235192}{{CMS Collaboration},
  ``Investigations of the impact of the parton shower tuning in {Pythia} 8 in
  the modelling of $\mathrm{t\overline{t}}$ at $\sqrt{s}=8$ and 13\,{TeV}'',}
  CMS Physics Analysis Summary CMS-PAS-TOP-16-021, 2016.

\bibitem{Czakon:2011xx}
\hrefCMSnoop {}{M.~Czakon and A.~Mitov, ``{Top++}: a program for the
  calculation of the top-pair cross-section at hadron colliders'',} \textit{
  Comput. Phys. Commun.} \textbf{ 185} (2014) 2930,
  \href{http://dx.doi.org/10.1016/j.cpc.2014.06.021}{\doi{10.1016/j.cpc.2014.06.021}},
\href{http://www.arXiv.org/abs/1112.5675}{\texttt{arXiv:1112.5675}}.

\bibitem{Botje:2011sn}
\hrefCMSnoop {}{M.~Botje { et~al.}, ``The {PDF4LHC} working group interim
  recommendations'',} 2011.
\href{http://www.arXiv.org/abs/1101.0538}{\texttt{arXiv:1101.0538}}.

\bibitem{Martin:2009bu}
\hrefCMSnoop {}{A.~D. Martin, W.~J. Stirling, R.~S. Thorne, and G.~Watt,
  ``Uncertainties on {$\alpS$} in global {PDF} analyses and implications for
  predicted hadronic cross sections'',} \textit{ Eur. Phys. J. C} \textbf{ 64}
  (2009) 653,
  \href{http://dx.doi.org/10.1140/epjc/s10052-009-1164-2}{\doi{10.1140/epjc/s10052-009-1164-2}},
\href{http://www.arXiv.org/abs/0905.3531}{\texttt{arXiv:0905.3531}}.

\bibitem{Gao:2013xoa}
J.~Gao\hrefCMSnoop {}{ { et~al.}, ``The {CT10} {NNLO} global analysis of
  {QCD}'',} \textit{ Phys. Rev. D} \textbf{ 89} (2014) 033009,
  \href{http://dx.doi.org/10.1103/PhysRevD.89.033009}{\doi{10.1103/PhysRevD.89.033009}},
\href{http://www.arXiv.org/abs/1302.6246}{\texttt{arXiv:1302.6246}}.

\bibitem{Ball:2013}
\hrefCMSnoop {}{{NNPDF} Collaboration, ``{Parton distributions with LHC
  data}'',} \textit{ Nucl. Phys. B} \textbf{ 867} (2013) 244,
  \href{http://dx.doi.org/10.1016/j.nuclphysb.2012.10.003}{\doi{10.1016/j.nuclphysb.2012.10.003}},
\href{http://www.arXiv.org/abs/1207.1303}{\texttt{arXiv:1207.1303}}.

\bibitem{Khachatryan:2016mnb}
\hrefCMSnoop {}{{CMS Collaboration}, ``Measurement of differential cross
  sections for top quark pair production using the $\text{lepton}+\text{jets}$
  final state in proton-proton collisions at 13\,{TeV}'',} \textit{ Phys. Rev.
  D} \textbf{ 95} (2017) 092001,
  \href{http://dx.doi.org/10.1103/PhysRevD.95.092001}{\doi{10.1103/PhysRevD.95.092001}},
\href{http://www.arXiv.org/abs/1610.04191}{\texttt{arXiv:1610.04191}}.

\bibitem{differentialXSdilep13TeV}
\hrefCMSnoop {}{{CMS Collaboration}, ``Measurement of normalized differential
  $\mathrm{t}\bar{\mathrm{t}}$~cross sections in the dilepton channel from
  pp~collisions at $\sqrt{s} = 13$\,{TeV}'',} \textit{ JHEP} \textbf{ 04}
  (2018) 060,
  \href{http://dx.doi.org/10.1007/JHEP04(2018)060}{\doi{10.1007/JHEP04(2018)060}},
\href{http://www.arXiv.org/abs/1708.07638}{\texttt{arXiv:1708.07638}}.

\bibitem{singletopNLO1}
M.~Aliev\hrefCMSnoop {}{ { et~al.}, ``{HATHOR} -- {HAdronic} {Top} and {Heavy}
  quarks {crOss} section {calculatoR}'',} \textit{ Comput. Phys. Commun.}
  \textbf{ 182} (2011) 1034,
  \href{http://dx.doi.org/10.1016/j.cpc.2010.12.040}{\doi{10.1016/j.cpc.2010.12.040}},
\href{http://www.arXiv.org/abs/1007.1327}{\texttt{arXiv:1007.1327}}.

\bibitem{singletopNLO2}
P.~Kant\hrefCMSnoop {}{ { et~al.}, ``{HATHOR} for single top-quark production:
  Updated predictions and uncertainty estimates for single top-quark production
  in hadronic collisions'',} \textit{ Comput. Phys. Commun.} \textbf{ 191}
  (2015) 74,
  \href{http://dx.doi.org/10.1016/j.cpc.2015.02.001}{\doi{10.1016/j.cpc.2015.02.001}},
\href{http://www.arXiv.org/abs/1406.4403}{\texttt{arXiv:1406.4403}}.

\bibitem{Kidonakis:2013zqa}
\hrefCMSnoop {}{N.~Kidonakis, ``Top quark production'',} in \textit{
  Proceedings, {Helmholtz} {International} {Summer} {School} on {Physics} of
  {Heavy} {Quarks} and {Hadrons} ({HQ} 2013)}, p.~139.
\newblock 2014.
\newblock \href{http://www.arXiv.org/abs/1311.0283}{\texttt{arXiv:1311.0283}}.
\newblock
\href{http://dx.doi.org/10.3204/DESY-PROC-2013-03/Kidonakis}{\doi{10.3204/DESY-PROC-2013-03/Kidonakis}}.

\bibitem{Alwall:2007fs}
J.~Alwall\hrefCMSnoop {}{ { et~al.}, ``Comparative study of various algorithms
  for the merging of parton showers and matrix elements in hadronic
  collisions'',} \textit{ Eur. Phys. J. C} \textbf{ 53} (2008) 473,
  \href{http://dx.doi.org/10.1140/epjc/s10052-007-0490-5}{\doi{10.1140/epjc/s10052-007-0490-5}},
\href{http://www.arXiv.org/abs/0706.2569}{\texttt{arXiv:0706.2569}}.

\bibitem{Melnikov:2006kv}
\hrefCMSnoop {}{K.~Melnikov and F.~Petriello, ``Electroweak gauge boson
  production at hadron colliders through {$O(\alpS^2)$}'',} \textit{ Phys. Rev.
  D} \textbf{ 74} (2006) 114017,
  \href{http://dx.doi.org/10.1103/PhysRevD.74.114017}{\doi{10.1103/PhysRevD.74.114017}},
\href{http://www.arXiv.org/abs/hep-ph/0609070}{\texttt{arXiv:hep-ph/0609070}}.

\bibitem{fewz2}
\hrefCMSnoop {}{Y.~Li and F.~Petriello, ``Combining {QCD} and electroweak
  corrections to dilepton production in {FEWZ}'',} \textit{ Phys. Rev. D}
  \textbf{ 86} (2012) 094034,
  \href{http://dx.doi.org/10.1103/PhysRevD.86.094034}{\doi{10.1103/PhysRevD.86.094034}},
\href{http://www.arXiv.org/abs/1208.5967}{\texttt{arXiv:1208.5967}}.

\bibitem{Frixione:2002ik}
\hrefCMSnoop {}{S.~Frixione and B.~R. Webber, ``Matching {NLO} {QCD}
  computations and parton shower simulations'',} \textit{ JHEP} \textbf{ 06}
  (2002) 029,
  \href{http://dx.doi.org/10.1088/1126-6708/2002/06/029}{\doi{10.1088/1126-6708/2002/06/029}},
\href{http://www.arXiv.org/abs/hep-ph/0204244}{\texttt{arXiv:hep-ph/0204244}}.

\bibitem{wwNNLO}
T.~Gehrmann\hrefCMSnoop {}{ { et~al.}, ``{$\mathrm{W}^+\mathrm{W}^-$}
  production at hadron colliders in {NNLO} {QCD}'',} \textit{ Phys. Rev. Lett.}
  \textbf{ 113} (2014) 212001,
  \href{http://dx.doi.org/10.1103/PhysRevLett.113.212001}{\doi{10.1103/PhysRevLett.113.212001}},
\href{http://www.arXiv.org/abs/1408.5243}{\texttt{arXiv:1408.5243}}.

\bibitem{mcfm}
\hrefCMSnoop {}{J.~M. Campbell and R.~K. Ellis, ``{MCFM} for the {Tevatron} and
  the {LHC}'',} \textit{ Nucl. Phys. Proc. Suppl.} \textbf{ 205-206} (2010) 10,
  \href{http://dx.doi.org/10.1016/j.nuclphysbps.2010.08.011}{\doi{10.1016/j.nuclphysbps.2010.08.011}},
\href{http://www.arXiv.org/abs/1007.3492}{\texttt{arXiv:1007.3492}}.

\bibitem{Agostinelli:2002hh}
\hrefCMSnoop {}{{GEANT4} Collaboration, ``{\GEANTfour}---a simulation
  toolkit'',} \textit{ Nucl. Instrum. Meth. A} \textbf{ 506} (2003) 250,
\href{http://dx.doi.org/10.1016/S0168-9002(03)01368-8}{\doi{10.1016/S0168-9002(03)01368-8}}.

\bibitem{Betchart:2013nba}
\hrefCMSnoop {}{B.~A. Betchart, R.~Demina, and A.~Harel, ``Analytic solutions
  for neutrino momenta in decay of top quarks'',} \textit{ Nucl. Instrum. Meth.
  A} \textbf{ 736} (2014) 169,
  \href{http://dx.doi.org/10.1016/j.nima.2013.10.039}{\doi{10.1016/j.nima.2013.10.039}},
\href{http://www.arXiv.org/abs/1305.1878}{\texttt{arXiv:1305.1878}}.

\bibitem{Loginov:2010zz}
\hrefCMSnoop {}{A.~Loginov, ``Strategies of data-driven estimations of
  $\mathrm{t}\bar{\mathrm{t}}$~backgrounds in {ATLAS}'',} \textit{ Nuovo Cim.
  C} \textbf{ 033} (2010) 175,
\href{http://dx.doi.org/10.1393/ncc/i2010-10657-2}{\doi{10.1393/ncc/i2010-10657-2}}.

\bibitem{CMS-TOP-17-014}
\hrefCMSnoop {}{{CMS Collaboration}, ``{Measurements of
  $\mathrm{t\overline{t}}$ differential cross sections in proton-proton
  collisions at $\sqrt{s} = 13$\,TeV using events containing two leptons}'',}
  \textit{ JHEP} \textbf{ 02} (2019) 149,
  \href{http://dx.doi.org/10.1007/JHEP02(2019)149}{\doi{10.1007/JHEP02(2019)149}},
\href{http://www.arXiv.org/abs/1811.06625}{\texttt{arXiv:1811.06625}}.

\bibitem{CMS-TOP-11-002}
\hrefCMSnoop {}{{CMS Collaboration}, ``{Measurement of the $\mathrm{t\bar{t}}$
  production cross section and the top quark mass in the dilepton channel in
  $\mathrm{pp}$ collisions at $\sqrt{s}=7$\,TeV}'',} \textit{ JHEP} \textbf{
  07} (2011) 049,
  \href{http://dx.doi.org/10.1007/JHEP07(2011)049}{\doi{10.1007/JHEP07(2011)049}},
\href{http://www.arXiv.org/abs/1105.5661}{\texttt{arXiv:1105.5661}}.

\bibitem{CMS-TOP-12-028}
\hrefCMSnoop {}{{CMS Collaboration}, ``Measurement of the differential cross
  section for top quark pair production in pp collisions at $\sqrt{s}$ =
  8\,{TeV}'',} \textit{ Eur. Phys. J. C} \textbf{ 75} (2015) 542,
  \href{http://dx.doi.org/10.1140/epjc/s10052-015-3709-x}{\doi{10.1140/epjc/s10052-015-3709-x}},
\href{http://www.arXiv.org/abs/1505.04480}{\texttt{arXiv:1505.04480}}.

\bibitem{Bernreuther:2004jv}
\hrefCMSnoop {}{W.~Bernreuther, A.~Brandenburg, Z.~G. Si, and P.~Uwer, ``Top
  quark pair production and decay at hadron colliders'',} \textit{ Nucl. Phys.
  B} \textbf{ 690} (2004) 81,
  \href{http://dx.doi.org/10.1016/j.nuclphysb.2004.04.019}{\doi{10.1016/j.nuclphysb.2004.04.019}},
\href{http://www.arXiv.org/abs/hep-ph/0403035}{\texttt{arXiv:hep-ph/0403035}}.

\bibitem{Khachatryan:2015hba}
\hrefCMSnoop {}{{CMS Collaboration}, ``Measurement of the top quark mass using
  proton-proton data at $\sqrt{s} = 7$ and 8\,{TeV}'',} \textit{ Phys. Rev. D}
  \textbf{ 93} (2016) 072004,
  \href{http://dx.doi.org/10.1103/PhysRevD.93.072004}{\doi{10.1103/PhysRevD.93.072004}},
\href{http://www.arXiv.org/abs/1509.04044}{\texttt{arXiv:1509.04044}}.

\bibitem{Aaboud:2016ymp}
\hrefCMSnoop {}{{ATLAS Collaboration}, ``Measurement of the inclusive
  cross-sections of single top-quark and top-antiquark $t$-channel production
  in pp~collisions at $\sqrt{s} = 13$\,{TeV} with the {ATLAS} detector'',}
  \textit{ JHEP} \textbf{ 04} (2017) 086,
  \href{http://dx.doi.org/10.1007/JHEP04(2017)086}{\doi{10.1007/JHEP04(2017)086}},
\href{http://www.arXiv.org/abs/1609.03920}{\texttt{arXiv:1609.03920}}.

\bibitem{Sirunyan:2018rlu}
\hrefCMSnoop {}{{CMS Collaboration}, ``Measurement of the single top quark and
  antiquark production cross sections in the $t$ channel and their ratio in
  proton-proton collisions at $\sqrt{s}= 13$\,{TeV}'',} 2018.
  \href{http://www.arXiv.org/abs/1812.10514}{\texttt{arXiv:1812.10514}}.
Submitted to \textit{Phys. Lett. B}.

\bibitem{Sirunyan:2018lcp}
\hrefCMSnoop {}{{CMS Collaboration}, ``Measurement of the production cross
  section for single top quarks in association with {W}~bosons in proton-proton
  collisions at $\sqrt{s}=13$\,{TeV}'',} \textit{ JHEP} \textbf{ 10} (2018)
  117,
  \href{http://dx.doi.org/10.1007/JHEP10(2018)117}{\doi{10.1007/JHEP10(2018)117}},
\href{http://www.arXiv.org/abs/1805.07399}{\texttt{arXiv:1805.07399}}.

\bibitem{Sirunyan:2017uzs}
\hrefCMSnoop {}{{CMS Collaboration}, ``Measurement of the cross section for top
  quark pair production in association with a {W} or {Z}~boson in proton-proton
  collisions at $\sqrt{s} = 13$\,{TeV}'',} \textit{ JHEP} \textbf{ 08} (2018)
  011,
  \href{http://dx.doi.org/10.1007/JHEP08(2018)011}{\doi{10.1007/JHEP08(2018)011}},
\href{http://www.arXiv.org/abs/1711.02547}{\texttt{arXiv:1711.02547}}.

\bibitem{Aaboud:2019njj}
\hrefCMSnoop {}{{ATLAS Collaboration}, ``Measurement of the
  {$\mathrm{t\bar{t}Z}$} and {$\mathrm{t\bar{t}W}$} cross sections in
  proton-proton collisions at $\sqrt{s}=13$\,{TeV} with the {ATLAS}
  detector'',} \textit{ Phys. Rev. D} \textbf{ 99} (2019) 072009,
  \href{http://dx.doi.org/10.1103/PhysRevD.99.072009}{\doi{10.1103/PhysRevD.99.072009}},
\href{http://www.arXiv.org/abs/1901.03584}{\texttt{arXiv:1901.03584}}.

\bibitem{barlowbeeston}
\hrefCMSnoop {}{R.~Barlow and C.~Beeston, ``Fitting using finite {M}onte
  {C}arlo samples'',} \textit{ Comput. Phys. Commun.} \textbf{ 77} (1993) 219,
\href{http://dx.doi.org/10.1016/0010-4655(93)90005-W}{\doi{10.1016/0010-4655(93)90005-W}}.

\bibitem{Cleveland79}
\hrefCMSnoop {}{W.~S. Cleveland, ``Robust locally weighted regression and
  smoothing scatterplots'',} \textit{ J. Am. Stat. Assoc.} \textbf{ 74} (1979)
  829, \href{http://dx.doi.org/10.2307/2286407}{\doi{10.2307/2286407}}.

\bibitem{Cleveland88}
\hrefCMSnoop {}{W.~S. Cleveland and S.~J. Devlin, ``Locally-weighted
  regression: {A}n approach to regression analysis by local fitting'',}
  \textit{ J. Am. Stat. Assoc.} \textbf{ 83} (1988) 596,
  \href{http://dx.doi.org/10.2307/2289282}{\doi{10.2307/2289282}}.

\bibitem{Baker:1983tu}
\hrefCMSnoop {}{S.~Baker and R.~D. Cousins, ``Clarification of the use of
  $\chi^2$ and likelihood functions in fits to histograms'',} \textit{ Nucl.
  Instrum. Meth.} \textbf{ 221} (1984) 437,
\href{http://dx.doi.org/10.1016/0167-5087(84)90016-4}{\doi{10.1016/0167-5087(84)90016-4}}.

\bibitem{Cowan:2010js}
\hrefCMSnoop {}{G.~Cowan, K.~Cranmer, E.~Gross, and O.~Vitells, ``Asymptotic
  formulae for likelihood-based tests of new physics'',} \textit{ Eur. Phys. J.
  C} \textbf{ 71} (2011) 1554,
  \href{http://dx.doi.org/10.1140/epjc/s10052-011-1554-0}{\doi{10.1140/epjc/s10052-011-1554-0}},
  \href{http://www.arXiv.org/abs/1007.1727}{\texttt{arXiv:1007.1727}}.
[Erratum: \emph{Eur. Phys. J. C} \textbf{73} (2013) 2501,
  \DOI{10.1140/epjc/s10052-013-2501-z}].

\bibitem{CMS-NOTE-2011-005}
\href {https://cds.cern.ch/record/1379837}{{ATLAS and CMS Collaborations, LHC
  Higgs Combination Group}, ``Procedure for the {LHC} {Higgs} boson search
  combination in summer 2011'',} Technical Report CMS-NOTE-2011-005,
  ATL-PHYS-PUB-2011-11, 2011.

\bibitem{Junk:1999kv}
\hrefCMSnoop {}{T.~Junk, ``Confidence level computation for combining searches
  with small statistics'',} \textit{ Nucl. Instrum. Meth. A} \textbf{ 434}
  (1999) 435,
  \href{http://dx.doi.org/10.1016/S0168-9002(99)00498-2}{\doi{10.1016/S0168-9002(99)00498-2}},
\href{http://www.arXiv.org/abs/hep-ex/9902006}{\texttt{arXiv:hep-ex/9902006}}.

\bibitem{Read:2002hq}
\hrefCMSnoop {}{A.~L. Read, ``Presentation of search results: the {\CLs}
  technique'',} \textit{ J. Phys. G} \textbf{ 28} (2002) 2693,
\href{http://dx.doi.org/10.1088/0954-3899/28/10/313}{\doi{10.1088/0954-3899/28/10/313}}.

\bibitem{RooMomentMorph}
\hrefCMSnoop {}{M.~Baak, S.~Gadatsch, R.~Harrington, and W.~Verkerke,
  ``Interpolation between multi-dimensional histograms using a new non-linear
  moment morphing method'',} \textit{ Nucl. Instrum. Meth. A} \textbf{ 771}
  (2015) 39,
  \href{http://dx.doi.org/10.1016/j.nima.2014.10.033}{\doi{10.1016/j.nima.2014.10.033}},
\href{http://www.arXiv.org/abs/1410.7388}{\texttt{arXiv:1410.7388}}.

\bibitem{Demortier:1099967}
\hrefCMSnoop {}{L.~Demortier, ``{$P$} values and nuisance parameters'',} in
  \textit{ Statistical issues for {LHC} physics. {Proceedings, Workshop,
  PHYSTAT-LHC, Geneva, Switzerland, June} 27-29, 2007}, p.~23.
\newblock 2008.
\newblock
\href{http://dx.doi.org/10.5170/CERN-2008-001}{\doi{10.5170/CERN-2008-001}}.

\bibitem{Kuhn:2013zoa}
\hrefCMSnoop {}{J.~H. K{\"{u}}hn, A.~Scharf, and P.~Uwer, ``Weak interactions
  in top-quark pair production at hadron colliders: {A}n update'',} \textit{
  Phys. Rev. D} \textbf{ 91} (2015) 014020,
  \href{http://dx.doi.org/10.1103/PhysRevD.91.014020}{\doi{10.1103/PhysRevD.91.014020}},
\href{http://www.arXiv.org/abs/1305.5773}{\texttt{arXiv:1305.5773}}.

\end{thebibliography}\endgroup
\cleardoublepage \appendix\section{The CMS Collaboration \label{app:collab}}\begin{sloppypar}\hyphenpenalty=5000\widowpenalty=500\clubpenalty=5000
\vskip\cmsinstskip
\textbf{Yerevan Physics Institute, Yerevan, Armenia}\\*[0pt]
A.M.~Sirunyan$^{\textrm{\dag}}$, A.~Tumasyan
\vskip\cmsinstskip
\textbf{Institut f\"{u}r Hochenergiephysik, Wien, Austria}\\*[0pt]
W.~Adam, F.~Ambrogi, T.~Bergauer, J.~Brandstetter, M.~Dragicevic, J.~Er\"{o}, A.~Escalante~Del~Valle, M.~Flechl, R.~Fr\"{u}hwirth\cmsAuthorMark{1}, M.~Jeitler\cmsAuthorMark{1}, N.~Krammer, I.~Kr\"{a}tschmer, D.~Liko, T.~Madlener, I.~Mikulec, N.~Rad, J.~Schieck\cmsAuthorMark{1}, R.~Sch\"{o}fbeck, M.~Spanring, D.~Spitzbart, W.~Waltenberger, C.-E.~Wulz\cmsAuthorMark{1}, M.~Zarucki
% ...
\vskip\cmsinstskip
\dag: Deceased\\
1:  Also at Vienna University of Technology, Vienna, Austria\\
% ...
\end{sloppypar}
\end{document}